\documentclass{article}

\usepackage[final]{neurips_2022}

\usepackage[utf8]{inputenc} 
\usepackage[T1]{fontenc}    
\usepackage{hyperref}       
\usepackage{url}            
\usepackage{booktabs}       
\usepackage{amsfonts}       
\usepackage{nicefrac}       
\usepackage{microtype}      
\usepackage{xcolor}         
\usepackage{graphicx}

\usepackage{amsmath}
\usepackage{amssymb}
\usepackage{mathtools}
\usepackage{amsthm}

\usepackage{algorithm}
\usepackage{algorithmic}
\usepackage{color,soul}
\usepackage{amsfonts}
\usepackage{amsmath}
\usepackage{tabularx}
\usepackage{multirow}
\usepackage{xspace}
\usepackage{booktabs}
\usepackage{wrapfig}

\usepackage[capitalize,noabbrev]{cleveref}

\theoremstyle{plain}

\theoremstyle{definition}

\theoremstyle{remark}

\newcommand{\review}[1]{{\color{black}{#1}}}

\def\eg{e.g.,\xspace} 
\def\ie{i.e,\xspace} 
\title{ELIGN: Expectation Alignment \\ as a Multi-Agent Intrinsic Reward}

\author{%
  Zixian Ma$^1$, Rose Wang$^1$, Li Fei-Fei$^1$, Michael Bernstein$^1$, Ranjay Krishna$^{1,2}$ \\
  Stanford University$^1$, University of Washington$^2$\\
  \texttt{\{zixianma,rewang,feifeili,msb,ranjaykrishna\}@cs.stanford.edu} \\
}

\begin{document}
\maketitle

\begin{abstract}
Modern multi-agent reinforcement learning frameworks rely on centralized training and reward shaping to perform well.
However, centralized training and dense rewards are not readily available in the real world. 
Current multi-agent algorithms struggle to learn in the alternative setup of decentralized training or sparse rewards.
To address these issues, we propose a self-supervised intrinsic reward  \textit{ELIGN - expectation alignment - } inspired by the self-organization principle in Zoology.
Similar to how animals collaborate in a decentralized manner with those in their vicinity, agents trained with expectation alignment learn behaviors that match their neighbors' expectations.
This allows the agents to learn collaborative behaviors without any external reward or centralized training.
We demonstrate the efficacy of our approach across 6 tasks in the multi-agent particle and the complex Google Research football environments, comparing ELIGN to sparse and curiosity-based intrinsic rewards.
When the number of agents increases, ELIGN scales well in all multi-agent tasks except for one where agents have different capabilities.
We show that agent coordination improves through expectation alignment because agents learn to divide tasks amongst themselves, break coordination symmetries, and confuse adversaries.
These results identify tasks where expectation alignment is a more useful strategy than curiosity-driven exploration for multi-agent coordination, enabling agents to do zero-shot coordination.

\end{abstract}

\section{Introduction}
\label{sec:introduction}
Many real world AI applications can be formulated as multi-agent systems, including autonomous vehicles~\citep{cao2012overview}, resource management~\citep{ying2005multi}, traffic control~\citep{sunehag2017value}, robot swarms~\citep{swamy2020scaled}, and multi-player video games~\citep{berner2019dota}.
Agents must adapt their behaviors to each other in order to coordinate successfully in these systems.
However, adaptive coordination algorithms are challenging to develop because each agent is not privy to other agents' intentions and their future behaviors~\citep{foerster2017learning}.

These challenges are more acute in decentralized training under partial observability than centralized training or full observability.
In the real world, agents act under partial observability and learn in a decentralized manner:
they do not learn collaborative behaviors with a single centralized algorithm with a complete knowledge of the environment~\citep{iqbal2019actor,liu2020pic}. 
Unfortunately, the most successful multi-agent algorithms train agents with a centralized critic, assuming access to all agents' observations and actions~\citep{foerster2018counterfactual,rashid2018qmix,sunehag2017value,lowe2017multi}.
The most successful multi-agent algorithms for decentralized training and partial observability assume task-specific reward shaping~\citep{jain2020cordial,iqbal2019actor}, which is expensive to generate.
These algorithms struggle to learn with sparse reward structure.

Consider a cooperative navigation task, where $N$ agents aim to simultaneously occupy $N$ goal locations. A centralized algorithm with full observability is capable of optimally assigning the nearest goal location to each respective agent. However, with partial observability, agents can see only a handful of goal locations and other agents. They are unaware of others' observations, actions, and intentions with decentralized training. We observe that agents simultaneously occupy the same goal; they fail to collaborate because they do not predict which goal each agent is expected to occupy.
To overcome instances of miscoordination, decentralized algorithms have adapted single-agent curiosity-based intrinsic rewards~\citep{pathak2017curiosity,stadie2015incentivizing}. Multi-agent curiosity-based rewards incentivize agents to explore novel states~\citep{iqbal2020coordinated}. Although curiosity helps agents discover new goal locations, it doesn't solve the challenge of coordination, such as assigning goals to each agent. Only a few attempts explore other forms of multi-agent intrinsic rewards~\review{\citep{iqbal2020coordinated,bohmer2019exploration,schafer2019curiosity}.}

\begin{figure*}
    \centering
    \includegraphics[width=\linewidth]{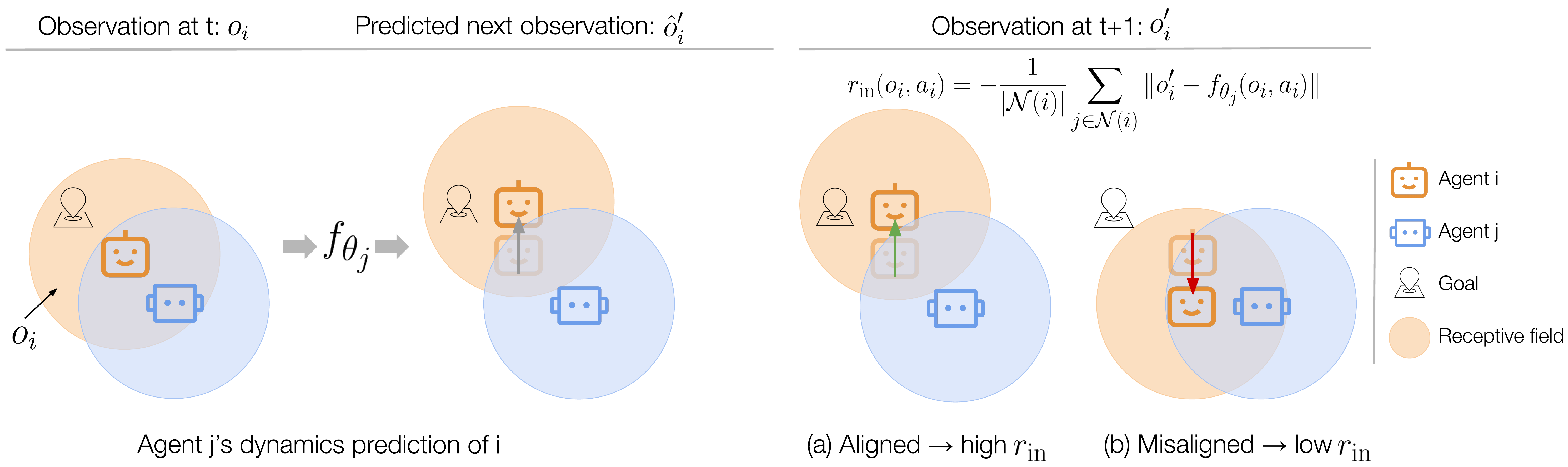}
    \caption{We introduce ELIGN, \ie expectation alignment, a task-agnostic intrinsic reward to improve multi-agent systems. Intuitively, ELIGN encourages agents to become more predictable to their neighbors. An agent (\eg agent $i$ here) learns to behave in ways that match its neighbors' (\eg agent $j$'s) predictions of its next observation. Here, agent $j$ expects agent $i$ to move up instead of down, moving closer to a point of interest above it. Agent $i$ attains (a) a higher reward when its action (\eg upward) aligns with this expectation or (b) a lower reward when its action (\eg downward) is misaligned.}
    \label{fig:alignment}
\end{figure*}

In this work, we propose ELIGN as a novel multi-agent self-supervised intrinsic reward, enabling decentralized training under partial observability. 
Intuitively, expectation alignment encourages agents to elicit behaviors that decrease future uncertainty for their team: it encourages each agent to choose actions that match their teammates' expectations.  
Going back to the cooperative navigation task, expectation alignment encourages each agent to move to goals others expect it to occupy, like goals that are either closest to the agent or goals that other agents aren't moving towards (Figure~\ref{fig:alignment}).
We take inspiration from the self-organization principle in Zoology~\citep{couzin2007collective}.
This principle hypothesizes that collective animal intelligence emerges because groups synchronize their behaviors using only their local environment; they do not rely on complete information about other agents and can coordinate successfully by predicting the dynamics of agents within their field-of-view~\citep{collett1998spatial,theraulaz1995coordination,ben1994generic,buhl2006disorder}.
Similarly, expectation alignment as an intrinsic reward is calculated based on the agent's local observations and its approximation of neighboring agents' expectations.
It does not require a centralized controller nor full observability.
ELIGN is task-agnostic and we apply it to both collaborative and competitive multi-agent tasks.

We demonstrate the efficacy of our approach in the multi-agent particle and Google Research football environments, two popular benchmarks for multi-agent reinforcement learning~\citep{lowe2017multi,kurach2019gfootball}.
We evaluate ELIGN under partial and full observability, with decentralized and centralized training, and in terms of scalability.
We observe that expectation alignment outperforms sparse and curiosity-based intrinsic rewards~\citep{ndousse2021emergent,stadie2015incentivizing,iqbal2020coordinated}, especially under partial observability with decentralized training.
We additionally test expectation alignment as a way to perform zero-shot coordination with new agent partners, and investigate why ELIGN improves coordination.
We show that agent coordination improves through expectation alignment because agents learn to divide tasks amongst themselves and break coordination symmetries~\citep{hu2020other}. 


\section{Related Work} 

Our formulation of expectation alignment, a task-agnostic intrinsic reward for multi-agent training, draws inspiration from the self-organization principle in Zoology, which posits that synchronized group behavior is mediated by local behavioral rules~\citep{couzin2007collective} and not by a centralized controller~\citep{camazine2020self}. Group cohesion emerges by predicting and adjusting one's behavior to that of near neighbors~\citep{buhl2006disorder}. This principle underlies the coordination found in multi-cellular organisms~\citep{camazine2020self}, the migration of wingless locusts~\citep{collett1998spatial}, the collective swarms of bacteria~\citep{ben1994generic}, the construction of bridge structures by ants~\citep{theraulaz1995coordination}, and some human navigation behaviors~\citep{couzin2007collective}.

\noindent\textbf{Intrinsic motivation for single agents.}
Although we draw inspiration from Zoology for formalizing expectation alignment as an intrinsic reward, there is a rich body of work on intrinsic rewards within the single-agent reinforcement learning community. 
To incentivize exploration, even when non-optimal successful trajectories are uncovered first, scholars have argued for the use of intrinsic motivation~\citep{schmidhuber1991possibility}. Single-agent intrinsic motivation has focused on exploring previously unencountered states~\citep{pathak2017curiosity,burda2018large}, which works particularly well in discrete domains. In continuous domains, identifying unseen states requires keeping track of an intractable number of visited states; instead, literature has recommended learning a forward dynamics model to predict future states and identify novel states using the uncertainty of this model~\citep{achiam2017surprise}. Other formulations encourage re-visiting states where the dynamics model's prediction of future states errs~\citep{stadie2015incentivizing,pathak2017curiosity}.
Follow up papers have improved how uncertainty~\citep{kim2020active} and model errors~\citep{burda2018exploration,sekar2020planning} are calculated.

\noindent\textbf{Intrinsic motivation for multiple agents.}
Most multi-agent intrinsic rewards have been adapted from single-agent curiosity-based incentives~\citep{iqbal2019coordinated,bohmer2019exploration,schafer2019curiosity} and have primarily focused on cooperative tasks. 
They propose intrinsic rewards to improve either coordination, collaboration, or deception: 
These rewards either maximize information conveyed by an agent's actions~\citep{iqbal2019coordinated,chitnis2020intrinsic,wang2019influence}, 
shape the influence of an agent~\citep{jaques2019social,foerster2017learning}, 
incentivize agents to hide intentions~\citep{strouse2018learning}, 
build accurate models of other agents' policies~\citep{hernandez2019agent,jaques2019social}, 
or break extrinsic rewards for better credit assignment~\citep{du2019liir}. 

Several multi-agent intrinsic rewards~\citep{hernandez2019agent,jaques2019social}, including ours, rely on the ability to model others' dynamics in a shared environment. This ability is a key component to coordination, closely related to Theory of Mind~\citep{tomasello2005understanding}. 
Our work can be interpreted as using a Theory of Mind model of others' behaviors to calculate an intrinsic motivation loss. Unlike existing Theory of Mind methods that learn a model per collaborator~\citep{roy2020promoting}, we learn a single dynamics model, allowing our method to scale as the number of agents increase.
Our proposal is related to model-based reinforcement learning~\citep{jaderberg2016reinforcement,wang2020modelbased}; however, instead of learning a dynamics model for control, we learn a dynamics model as a source of reward.
Our work is closely related to a recently proposed auxiliary loss on predicting an agent's own future states~\citep{ndousse2021emergent}. 
However, there are three key differences. 
First, their work predicts ego-agent observations, whereas our work additionally predicts future observations from the other agents' point of view. 
Second, their loss optimizes state embeddings while ours optimizes agents' policies. 
Third, their work focuses on cooperative tasks whereas ours applies to both cooperative and competitive domains.

\noindent\textbf{Multi-agent reinforcement learning algorithms.}
Today, the predominant deep multi-agent framework uses actor-critic methods with a centralized critic and decentralized execution~\citep{lowe2017multi,foerster2018counterfactual,iqbal2019actor,liu2020pic,rashid2018qmix}. This framework allows a critic to access the observations and actions of all agents to ease training. 
However, there are several situations where centralized training may not be desirable or possible. 
Examples include low bandwidth communication restrictions or human-robot tasks where observations cannot be easily shared between agents~\citep{ying2005multi,cao2012overview,huang2015adaptive}. 
Decentralized training is therefore the most practical training paradigm but it suffers from unstable training: the environment is nonstationary from a single-agent's perspective~\citep{lowe2017multi}.
Our work uses a decentralized training framework and tackles the nonstationarity challenge with an intrinsic reward designed to improve an agent's ability to model others. 
We also apply expectation alignment to centralized training and observe that it still aids cooperative and some competitive tasks.

\section{Background}

We formulate our setting as a partially observable Markov game $(\mathcal{S}, \mathcal{O}, \mathcal{A}, \mathcal{T}, r_\textrm{ex}, N)$~\citep{littman1994markov}. A Markov game for $N$ agents is defined by a state space $\mathcal{S}$ describing the possible configurations of the environment. 
The observation space for agents is $\mathcal{O} = (\mathcal{O}_1, \ldots, \mathcal{O}_N)$ and the action space is $\mathcal{A} = (\mathcal{A}_1, \ldots, \mathcal{A}_N)$.
Each agent $i$ observes $\mathbf{o}_i \in \mathcal{O}_i$, a private partial view of the state, and performs actions $a_i \in \mathcal{A}_i$. 
Using the observation, each agent uses a stochastic policy $\pi_{\theta_i}: \mathcal{O}_i \times \mathcal{A}_i \rightarrow [0, 1]$, where $\theta_i$ parameterizes the policy. 
The environment changes according to the state transition function which transitions to the next state using the current state and each agent's actions, $\mathcal{T}: \mathcal{S} \times \mathcal{A} \rightarrow \mathcal{S}$.
The team of agents obtains a shared extrinsic reward as a function of the environment state, $r_\textrm{ex}: \mathcal{S} \times \mathcal{A} \rightarrow \mathbb{R}$. The team's goal is to maximize the total expected return: $R = \sum_{t=0}^{T} \gamma^t r_\textrm{ex}^t$ where $0 \leq \gamma \leq 1$ is the discount factor, $t$ is the time step, and $T$ is the time horizon.  
The environment may also contain adversarial agents who have their own reward structure.

\section{Expectation Alignment}
\label{sec:model}
To understand expectation alignment intuitively, let's revisit the cooperative navigation task, where $N$ agents are rewarded for simultaneously occupying as many goal locations as possible. 
In Figure~\ref{fig:alignment}, agent $i$ has a dynamics model trained on its past experiences. 
It predicts how future states will evolve from the point of view of agent $j$, who is within $i$'s view. 
In this example, $j$ will expect $i$ to move towards the goal since $i$ is closer to it. ELIGN encourages $i$ to pursue the action that $j$ expects (Figure~\ref{fig:alignment}(a)). 
In turn, $j$ can now assume that the observed goal location will eventually be occupied by $i$ and should therefore explore to find another goal.
By aligning shared expectations, agent behaviors become more predictable.
Conversely, when neighbors behave opposite to an agent's predictions, the agent can infer about the environment outside of its own receptive field~\citep{krause2002living}. 
For example, in Figure~\ref{fig:alignment} (b), if agent $j$ observes $i$ running away from a goal, this surprising behavior might indicate the existence of an adversary outside $j$'s receptive field.

Our training algorithm consists of three interwoven phases of learning a dynamics model, calculating the ELIGN reward, and optimizing the agent's policy (Algorithm~\ref{alg:alignment}). 

\begin{wrapfigure}[24]{R}{0.48\textwidth}
\vspace{-3\intextsep}
    \begin{minipage}{0.5\textwidth}
      \begin{algorithm}[H]
        \begin{algorithmic}[1]
    \STATE Initialize replay buffer $D$ and $D^\prime$\\
    \STATE Initialize $N$ agents with random $\theta_i$: $i \in [1, N]$\\
    \WHILE{not converged}
        \FOR{$b = 1\ldots B$}
            \STATE Populate buffer $D$ with episode using policies $(\pi_{\theta_1}, \ldots, \pi_{\theta_N})$ \\
        \ENDFOR
        \STATE // \textsc{Train dynamics model} \\
        \FOR{agent $i = 1 \ldots N$} 
            \STATE Sample transitions: $\{(o_i , a_i, r_\textrm{ex}, o^\prime_i)\} \sim D_i$  \\
            \STATE Predict $\hat{o}_i^\prime = f_{\theta_i}(o_i, a_i)$\\
            \STATE Update dynamics $\theta_i$ using $o^\prime_i$. \\
        \ENDFOR
        \STATE // \textsc{Calculate ELIGN reward} \\
        \FOR{agent $i = 1 \ldots N$}
            \STATE Sample $B$ transitions: $\{(o_i, a_i, r_\textrm{ex}, o^\prime_i)\} \sim D_i$  \\
            \STATE Compute intrinsic rewards $r_\textrm{in}(o_i, a_i)$
            \STATE Add $\{(o_i, a_i, r_\textrm{ex}+\beta r_\textrm{in}, o^\prime_i)\}$ to $D^\prime_i$
        \ENDFOR
        \STATE // \textsc{Policy learning} \\
        \STATE Update all $\theta_i$s using transitions from $D^\prime$ 
    \ENDWHILE
\end{algorithmic}
\caption{ELIGN: Expectation Alignment}
\label{alg:alignment}
      \end{algorithm}
    \end{minipage}
  \end{wrapfigure}

\subsection{Training the dynamics model}
Similar to prior work~\citep{wang2018look,kidambi2020morel}, each agent $i$ learns a dynamics model $f_{\theta_i}$ to predict the next observation $\hat{o}^\prime_i$ given its current observation and action $o_i, a_i$, \ie
$$\hat{o}^\prime_i = f_{\theta_i}(o_i, a_i).$$ 
We use a three-layer Multi-Layer Perceptron with ReLU non-linearities as the dynamics model. We minimize the mean squared error between its prediction and ground truth next observation $o^\prime_i$.

\subsection{Calculating intrinsic reward}
The intrinsic reward captures how well agent $i$ aligns to its neighbors' (\eg agent $j$'s) expectations on its next state. Calculating this reward requires $j$ to accurately predict $i$'s behavior, simulating a Theory of Mind~\citep{tomasello2005understanding}. 
As suggested by the self-organization principle, $i$ must learn to align to $j$'s predictions. Ideally, the ELIGN intrinsic reward is calculated as:
$$r_\textrm{in}(o_i, a_i) = - \frac{1}{|\mathcal{N}(i)|} \sum_{j\in \mathcal{N}(i)}{\|o^\prime_{i} - f_{\theta_j}(o_{i}, a_{i})\|}$$
where $\mathcal{N}(i)$ is the set of neighbors within $i$'s receptive field, including $i$ itself. The ELIGN reward is high when the average $L_2$ loss is small, \ie when $i$'s actual next observation is close to agent $j$'s predicted observation of $i$ for all $j$ in its neighbors.  In that case, $i$ has chosen an action that aligns with $j$'s expectations of how $i$ should act.

In a decentralized training setup, however, $i$ doesn't have access to $j$'s dynamics model $f_{\theta_j}$, so $i$ approximates $j$'s dynamic model with a proxy: its own dynamics model $f_{\theta_i}$ \review{and the knowledge of agent $j$'s observation radius}. Such an approximation is ecologically valid since we often approximate others' behaviors using a second-order cognitive Theory of Mind~\citep{morin2006levels}.
Additionally, $i$ doesn't have access to $j$'s entire observation; so, we restrict the future prediction from $j$'s point of view by using the portion of $j$'s observation $i$ can see: $o_{i \cap j} = o_i \odot o_j$. 
Agent $i$'s decentralized intrinsic reward then becomes:
$$r_\textrm{in}(o_i, a_i) = - \frac{1}{|\mathcal{N}(i)|} \sum_{j\in \mathcal{N}(i)}{\|o^\prime_{i \cap j} - f_{\theta_i}(o_{i \cap j}, a_{i})\|}$$
We found that the approximation of $f_{\theta_j}$ using $f_{\theta_i}$ works well \review{empirically}. Dynamics model losses for all agents quickly decrease within 5-10 training epochs. we validate its applicability in \review{small-scale} heterogeneous multi-agent tasks where agents have variable capabilities, \review{although we find the methods perform similarly when more heterogeneous agents are added.}


\subsection{Policy learning}
Once the ELIGN rewards are calculated, the total rewards at each step for each agent $i$ is: $r_i= r_\textrm{ex} + \beta r_\textrm{in}(o_i, a_i)$ where  $r_\textrm{ex}$ is the extrinsic reward provided by the environment and $\beta$ is a hyperparameter for weighing the intrinsic reward in the agent's overall reward calculation. In practice, we set $\beta$ to be $ \frac{1}{|\mathcal{O}_i|}$ where $|\mathcal{O}_i|$ is the observation dimension; we find this scale generalizes well across tasks.
Since our contribution is agnostic to any particular multi-agent training algorithm, the team of agents can now be trained using any multi-agent training algorithm to maximize returns $R = \sum_{t=0}^{T} \gamma^t r$. 

Both centralized and decentralized training algorithms can make use of these rewards. We primarily use the multi-agent decentralized variant of the soft-actor critic algorithm in our experiments~\citep{haarnoja2017soft,iqbal2019actor}.
Compared to centralized joint-action training, whose action space grows exponentially in $N$ agents, our decentralized method has linear space complexity. Further, decentralized training can parallelize training time to be less than linear with respect to $N$. Although we present results with one centralized training framework, studying the impact of expectation alignment with all the centralized critic frameworks is out of scope for this paper.

\subsection{Extending expectation alignment to competitive tasks}
We extend the ELIGN formulation to competitive tasks where a \review{team} of agents compete against  adversaries. In this case, agents are encouraged to \textit{misalign} with their adversaries' expectations, \ie agents are incentivized to be unpredictable to their adversaries within its receptive field ($\mathcal{N}_{\text{adv}}(i)$):
\begin{align*}
    r_{\text{in}} &= \frac{1}{|\mathcal{N}_{\text{adv}}(i)|} \sum_{k\in \mathcal{N}_{\text{adv}}(i)}{\|o^\prime_{i \cap k} - f_{\theta_i}(o_{i \cap k}, a_i))\|}
\end{align*}



\section{Experiments}
\label{sec:experiments}
Our experiments explore the utility of using expectation alignment as an intrinsic reward compared to sparse and curiosity-based intrinsic rewards. 
We primarily focus on decentralized training under partial observability. 
However, we also demonstrate that ELIGN can easily augment centralized methods and assist in fully observable tasks. 
We vary the number of agents in the multi-agent particle tasks to test scalability. 
We end by investigating how and why ELIGN improves coordination by designing three evaluation conditions. First, does expectation alignment improve coordination by breaking symmetries~\citep{hu2020other,wang2020cooks}? Second, does ELIGN enable zero-shot generalization to new partners? Lastly, is the dynamics model critical in aligning agent behaviors?

\subsection{Environments}
We evaluate ELIGN across both cooperative and competitive tasks in the multi-agent particle environment~\citep{mordatch2017emergence,lowe2017multi} and the Google Research football environment~\citep{kurach2019gfootball}. 

\noindent\textbf{State and action space}
The multi-agent particle environment is a two-dimensional world. The Google Research football environment is a three-dimensional world. Both environments have continuous state spaces and discrete action spaces. Particle agents observe all agents' positions and velocities. They can ``stay'' or change \review{their} velocity in one of the four cardinal directions. \review{Each football agent controls one player.} Players observe the ball, other players' positions and directions. They can apply one of ten actions from ``top\_left'', ``top'', ``top\_right'', ``right'', ``bottom\_right'', ``bottom'', ``bottom\_left'', ``sprint'', and ``dribble''.

\noindent\textbf{Observability}
The original environments assume full observability, where each agent can observe the position $p = (x, y)$ and velocity $v = (\Delta x, \Delta y)$ of all agents; each agent's observation vector is thus $\mathbf{o}_{i, \text{full}} = [p_1,\ldots,p_N, v_1, \ldots, v_N]$.
We extend these environments to be partially observable, where agent $i$ observes only the portion within its receptive field;
like prior work with partial observability~\citep{corder2019decentralized}, we hide the position and velocity information of any agent $j$ outside of agent $i$'s receptive field; \ie if the Euclidean distance between agent $i$ and $j$ surpasses a vicinity threshold $\tau$, then $p_j$ and $v_j$ are $0$ in $\mathbf{o}_{i, \text{partial}}$. 
We set $\tau=0.5$ for partially observable and $\infty$ in the original fully observable case, where the world's width and height are $2.0$ in the multi-agent particle environment and $0.84:2.00$ in the Google Research football environment. Both environments also contain objects such as obstacles, goals and a ball; they are similarly hidden if out of sight.
Partial observability is a more ecologically valid training condition since most agents in real-world tasks can only observe a small portion of their environment at a given time.

\begin{table*}[t]
\centering 
\caption{\label{tab:1_dec_reward}We report the mean test episode extrinsic rewards and standard errors of $\textit{decentralized training under partial observability}$ in multi-agent particle and Google Research football environments. $\textrm{\textsc{elign}}_\textrm{self/team}$ outperform \textsc{sparse} and both curiosity-based intrinsic rewards. $\textrm{\textsc{elign}}_\textrm{adv}$ achieves the best performance among all competitive tasks except for $\textit{Physical deception}$, where $\textrm{\textsc{elign}}_\textrm{team}$ is the best. These results demonstrate the benefit of using alignment as intrinsic reward to train better decentralized policies under partial observability.
}
\resizebox{\textwidth}{!}{
\begin{tabular}{lrp{0.001pt}rp{0.001pt} rp{0.001pt}rp{0.001pt}rp{0.001pt}r}
\multicolumn{1}{c}{}                & \multicolumn{3}{c}{\textbf{Cooperative}}                              & & \multicolumn{7}{c}{\textbf{Competitive}}                                                                            \\
\cmidrule{2-4}\cmidrule{6-12}
\multicolumn{1}{l}{Task (Agt\# v Adv\#) }                & \multicolumn{1}{c}{\textbf{Coop nav. (3v0)}} && \multicolumn{1}{c}{\textbf{Hetero nav. (4v0)}} && \multicolumn{1}{c}{\textbf{Phy decep. (2v1)}}  && \multicolumn{1}{c}{\textbf{Pred-prey (2v2)}} && \multicolumn{1}{c}{\textbf{Keep-away (2v2)}} &&  \multicolumn{1}{c}{\textbf{3v1 w/ keeper (3v2)}}\\
\cmidrule{1-1}\cmidrule{2-2}\cmidrule{4-4}\cmidrule{6-6}\cmidrule{8-8}\cmidrule{10-10}\cmidrule{12-12}
  $\textsc{sparse}^1$                & $139.07\pm 13.63$           && $284.42\pm 12.83$ && $93.60\pm 8.61$ && $-4.72\pm 2.4$ && $4.58\pm 3.27$ &&  $0.020\pm0.001$\\
                          $\textrm{\textsc{curio}}_\textrm{self}^2$        & $133.93\pm 7.66$            && $286.22\pm  9.97$             && $68.80\pm  7.93$                 && $-6.50\pm  2.18$             && $11.88\pm2.88$            && $0.024\pm0.004$\\
                          $\textrm{\textsc{curio}}_\textrm{team}^3$           & $125.42\pm11.95$           && $262.28\pm22.59$            && $85.31\pm 11.93$                && $-3.57\pm 1.75$              && $9.54\pm 5.04$ &&  $0.021\pm0.002$ \\
                          $\textrm{\textsc{elign}}_\textrm{self}$        & $\mathbf{155.88}\pm \mathbf{5.11} $          && $292.34\pm 9.24$             && $69.91\pm 4.51$                 && $-7.58\pm2.55$              && $12.84\pm 4.29$            && $0.003\pm0.018$ \\
                          $\textrm{\textsc{elign}}_\textrm{team}$   & $141.04\pm            8.04$            && $\mathbf{311.67}\pm            \mathbf{10.88}$           && $\mathbf{101.72}\pm \mathbf{6.31}$                 && $-7.69\pm 2.69$              && $2.96\pm 4.03$                       && $0.022\pm 0.001$\\
                          $\textrm{\textsc{elign}}_\textrm{adv}$      & ---         && ---               && $92.20\pm 4.23$                 && $\mathbf{-2.51}\pm \mathbf{1.70}$              && $\mathbf{19.46}\pm \mathbf{5.05}$ && $\mathbf{0.025}\pm\mathbf{0.001}$\\
                          \hline
\review{$\textsc{Hand-crafted}^1$}                & \review{$75.56	\pm 18.90$}           && \review{$228.48	\pm 18.88$} && \review{$94.25	\pm 14.75$} && \review{$\mathbf{-0.77}	\pm \mathbf{0.17}$} && \review{$\mathbf{52.14}	\pm \mathbf{3.11}$} &&  \review{$-$}\\
                        \multicolumn{10}{l}{$^1$~\cite{lowe2017multi,kurach2019gfootball},$^2$~\cite{stadie2015incentivizing},$^3$~\cite{iqbal2020coordinated}}  
\end{tabular}
}
\end{table*}

\begin{figure*}
    \centering
    \includegraphics[width=\linewidth]{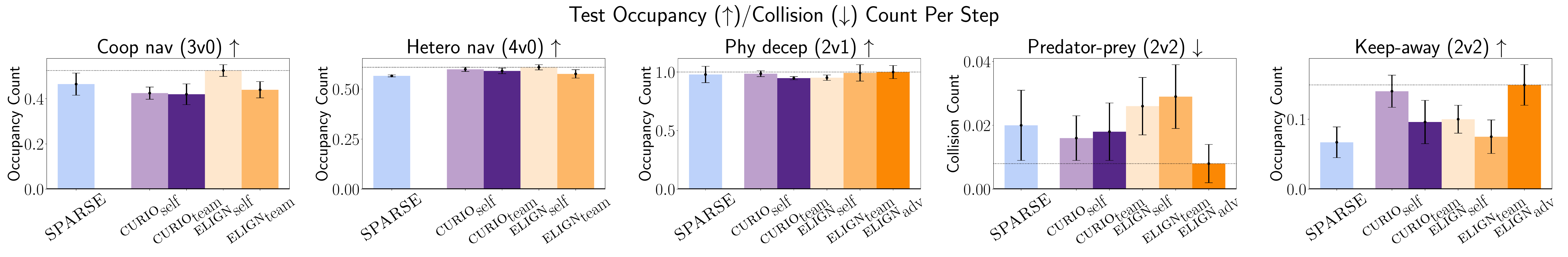}
    \vspace{-1em}
    \caption{We plot the average test occupancy/collision count per step of decentralized algorithms under partially observable multi-agent particle tasks. On these metrics, $\textrm{\textsc{elign}}_\textrm{self}$ and $\textrm{\textsc{elign}}_\textrm{adv}$ perform the best on cooperative and competitive tasks respectively.}
    \label{fig:dec_occ_small}
\end{figure*}

\subsection{Tasks}

\noindent\textbf{Multi-agent particle environment}
We use the following tasks from the multi-agent particle environment~\citep{lowe2017multi,liu2020pic}. 
We choose $N$ based on prior work. 

\noindent\textit{Cooperative navigation:} $N$ agents must cooperate to reach a set of $N$ goal locations. Agents are collectively rewarded based on the occupancy of any agent on any goal location.

\noindent\textit{Heterogeneous navigation:} $N$ agents must reach $N$ goals but they differ in speeds and sizes. $\frac{N}{2}$ agents are slow and big, and the other $\frac{N}{2}$ agents are fast and small. 

\noindent\textit{Physical deception:} 
$N$ agents cooperate to reach a single goal location and are rewarded if any one occupies the goal.
However, they are penalized when any of $\frac{N}{2}$ adversaries occupies the goal and gets rewarded. The adversaries do not know which landmark is the goal and must infer it from the agents' behavior. The agents should learn to deceive the adversaries by covering all the landmarks.

\noindent\textit{Keep-away:} There are $N$ landmarks, one of which is the goal 
and known to $N$ agents.
Agents are rewarded for occupying it and preventing $M$ adversaries from reaching it.
Adversaries are rewarded for pushing the agents away from the goal, but they can only infer which landmark is the goal.

\noindent\textit{Predator-prey:} $N$ slow adversaries chase and capture $N$ fast cooperating agents around a randomly generated obstacle-filled environment. Each time an adversary catches an agent, the agent is penalized and the adversary is rewarded.

\noindent\textbf{Google Research football}
We use the \textit{Academy 3vs1 with Keeper} competitive task from the Google Research football environment~\citep{kurach2019gfootball}. In this task, three agents try to score from the edge of the penalty box, one on each side, and the other at the center. This task is initialized with the centered agent having the ball and facing the defender. There is an adversary who plays the keeper.

\subsection{Training and evaluation}
We train all algorithms with $5$ random seeds. All the hyperparameters used in the training can be found in the Appendix.
For the Multi-agent particle environment,
each experiment uses one Tesla K40 GPU to train until convergence, i.e. the best evaluation episode reward hasn’t changed for $100$ epochs. Each epoch equates to $4K$ episodes of $25$ timesteps. We evaluate the algorithms by running $1K$ test episodes of $25$ timesteps and mainly report the mean average test episode reward and standard error across the random seeds. We also evaluate on task-specific metrics, including agent-goal occupancy/agent-adversary collision count, and agent-goal/agent-adversary distance. 
For Google Research football, each experiment trains for $5M$ timesteps. We evaluate on and report the mean average episode rewards and the standard errors across the seeds.

\subsection{Baselines}

\begin{table*}[t]
\centering 
\caption{\label{tab:3_scaled_dec_reward}We report the mean test episode extrinsic rewards and standard errors as the number of agents is increased and \review{trained} using of $\textit{decentralized}$ algorithms. When the number of agents increases, one of $\textrm{\textsc{elign}}$ still performs the best in all tasks except for \textit{Heterogenous navigation}.
}
\resizebox{\textwidth}{!}{
\begin{tabular}[width=\linewidth]{lrp{0.001pt}rp{0.001pt} rp{0.001pt}rp{0.001pt}r}
\multicolumn{1}{l}{}                & \multicolumn{3}{c}{\textbf{Cooperative}}                              & & \multicolumn{5}{c}{\textbf{Competitive}}                                                                            \\
\cmidrule{2-4}\cmidrule{6-10}
\multicolumn{1}{l}{Task (Agt \# vs. Adv \#) }                & \multicolumn{1}{c}{\textbf{Coop nav. (5v0)}} && \multicolumn{1}{c}{\textbf{Hetero nav. (6v0)}} && \multicolumn{1}{c}{\textbf{Phy decep. (4v2)}}  && \multicolumn{1}{c}{\textbf{Pred-prey (4v4)}} && \multicolumn{1}{c}{\textbf{Keep-away (4v4)}} \\
\cmidrule{1-1}\cmidrule{2-2}\cmidrule{4-4}\cmidrule{6-6}\cmidrule{8-8}\cmidrule{10-10}
 $\textsc{sparse}^1$             & $459.92\pm22.44$ && $616.62\pm25.30$ && $166.89\pm27.72$ && $-28.75\pm7.3$ && $0.75\pm1.82$        \\
                        $\textrm{\textsc{curio}}_\textrm{self}^2$        & $458.45\pm19.79$ && $\mathbf{702.73}\pm\mathbf{18.57}$ && $146.55\pm29.05$ && $-25.35\pm6.16$ && $10.52\pm5.48$            \\
                          $\textrm{\textsc{curio}}_\textrm{team}^3$           &$497.15\pm11.47$ && $695.38\pm12.22$ && $84.66\pm16.94$ && $-17.21\pm8.23$ && $1.40\pm2.06$\\
                                $\textrm{\textsc{elign}}_\textrm{self}$             & $\mathbf{498.24}\pm\mathbf{9.77}$ && $646.70\pm23.25$ && $137.38\pm30.00$ && $\mathbf{-9.14}\pm\mathbf{5.57}$ && $9.83\pm11.22$       \\
                                $\textrm{\textsc{elign}}_\textrm{team}$       & $488.83\pm20.82$ && $638.74\pm28.93$ && $\mathbf{186.83}\pm\mathbf{21.92}$ && $-20.4\pm5.93$ && $2.07\pm4.55$          \\
                                $\textrm{\textsc{elign}}_\textrm{adv}$          & --- && --- && $182.61\pm17.63$ && $-21.37\pm7.02$ && $\mathbf{11.29}\pm\mathbf{9.02}$       \\
                               \multicolumn{10}{l}{$^1$~\cite{lowe2017multi},$^2$~\cite{stadie2015incentivizing},$^3$~\cite{iqbal2020coordinated}}  
\end{tabular}
}
\end{table*}

All algorithms are trained using the same agent architecture and optimization algorithm. 
They vary in task-specific extrinsic rewards and intrinsic rewards.
We use two versions of the soft actor-critic algorithm~\citet{haarnoja2017soft}: a decentralized one that trains each agent individually without access to other agents' observations and actions (ie. the original soft-actor critic algorithm) and a centralized one with a critic that has access to other agents' observations and actions~\citep{iqbal2019actor}. 
Note, our intrinsic reward can also be added to non-actor-critic methods, such as COMA~\citep{foerster2018counterfactual} and VDN~\citep{sunehag2017value}.
We leave this to future work to avoid conflating the effects of expectation alignment with COMA's counterfactual reasoning and VDN's value decomposition. 

We use \textsc{sparse}~\citep{lowe2017multi,kurach2019gfootball}, $\textrm{\textsc{curio}}_\textrm{self}$~\citep{stadie2015incentivizing}, $\textrm{\textsc{curio}}_\textrm{team}$~\citep{iqbal2020coordinated}, and variations of our \textsc{elign} rewards.
\textsc{Sparse} rewards agents only when they reach \review{a} goal state. $\textrm{\textsc{curio}}_\textrm{team}$ is a curiosity-based multi-agent intrinsic reward which maximizes the average $L_2$ loss (instead of minimizing it in $\textsc{elign}$). It rewards agents for exploring novel states~\citep{iqbal2020coordinated}. 
$\textrm{\textsc{curio}}_\textrm{self}$ also maximizes the $L_2$ loss but only using agent $i$'s own observation~\citep{stadie2015incentivizing}. We experiment with three variants of ELIGN: $\textrm{\textsc{elign}}_\textrm{self}$, incentivizing alignment to one's own expectation $\textrm{\textsc{elign}}_\textrm{team}$ incentivizing agents to align to their team, and $\textrm{\textsc{elign}}_\textrm{adv}$ incentivizing misalignment to adversaries' expectations. 
Note that $\textrm{\textsc{elign}}_\textrm{self}$ is similar to the auxiliary loss in ~\citet{ndousse2021emergent} but we use it for policy optimization, rather than for training a state encoder. \review{We also add hand-crafted dense rewards to provide oracle performance for all tasks.}

\subsection{Results in partially observable environments with decentralized training}

\noindent\textbf{ELIGN outperforms baselines across cooperative and competitive tasks in the multi-agent particle environment.} Table~\ref{tab:1_dec_reward} demonstrates that both $\textrm{\textsc{elign}}_\textrm{self}$ and $\textrm{\textsc{elign}}_\textrm{team}$ outperform all \textsc{sparse} and $\textrm{\textsc{curio}}_\textrm{self/team}$ baseline rewards in cooperative tasks. While not all $\textrm{\textsc{elign}}$ variants surpass the baselines in competitive tasks, we find that $\textrm{\textsc{elign}}_\textrm{team}$ achieves the highest reward in \textit{Physical deception}, and $\textrm{\textsc{elign}}_\textrm{adv}$  performs the best in \textit{Predator-prey} and \textit{Keep-away}. Similarly, Figure~\ref{fig:dec_occ_small} shows that $\textrm{\textsc{elign}}_\textrm{self}$ achieves the highest per-step occupancy count in both cooperative tasks, and $\textrm{\textsc{elign}}_\textrm{adv}$ does the best in all competitive tasks.

\noindent\textbf{ELIGN outperforms baselines in the complex Google Research football environment.}
As shown in Table~\ref{tab:1_dec_reward}, $\textrm{\textsc{elign}}_\textrm{adv}$ achieves the best mean average episode reward in the competitive \textit{Academy 3vs1 with keeper} task. 
Collectively, these results provide empirical evidence that the self-organizing principle improves coordination under partial information, a setting that is most realistic to real world multi-agent systems.

\noindent\textbf{In competitive tasks, agents benefit more from being misaligned to adversaries than being aligned to their team members.} Among the four competitive tasks in the multi-agent particle and football environments, we find that $\textrm{\textsc{elign}}_\textrm{adv}$ outperforms all \textsc{sparse} and $\textrm{\textsc{curio}}_\textrm{self/team}$ baselines and other variants of $\textrm{\textsc{elign}}$ in \textit{Predator-prey}, \textit{Keep-away} and \textit{Academy 3vs1 with keeper}. This suggests that being misaligned to adversaries, ie taking surprising actions that conflict with the adversary's expectations, might be a more useful strategy in multi-agent competitive tasks.

\begin{wrapfigure}[12]{R}{0.3\linewidth}
    \centering
    \vspace{-1\intextsep}
    \hspace*{-.75\columnsep}\includegraphics[width=\linewidth]{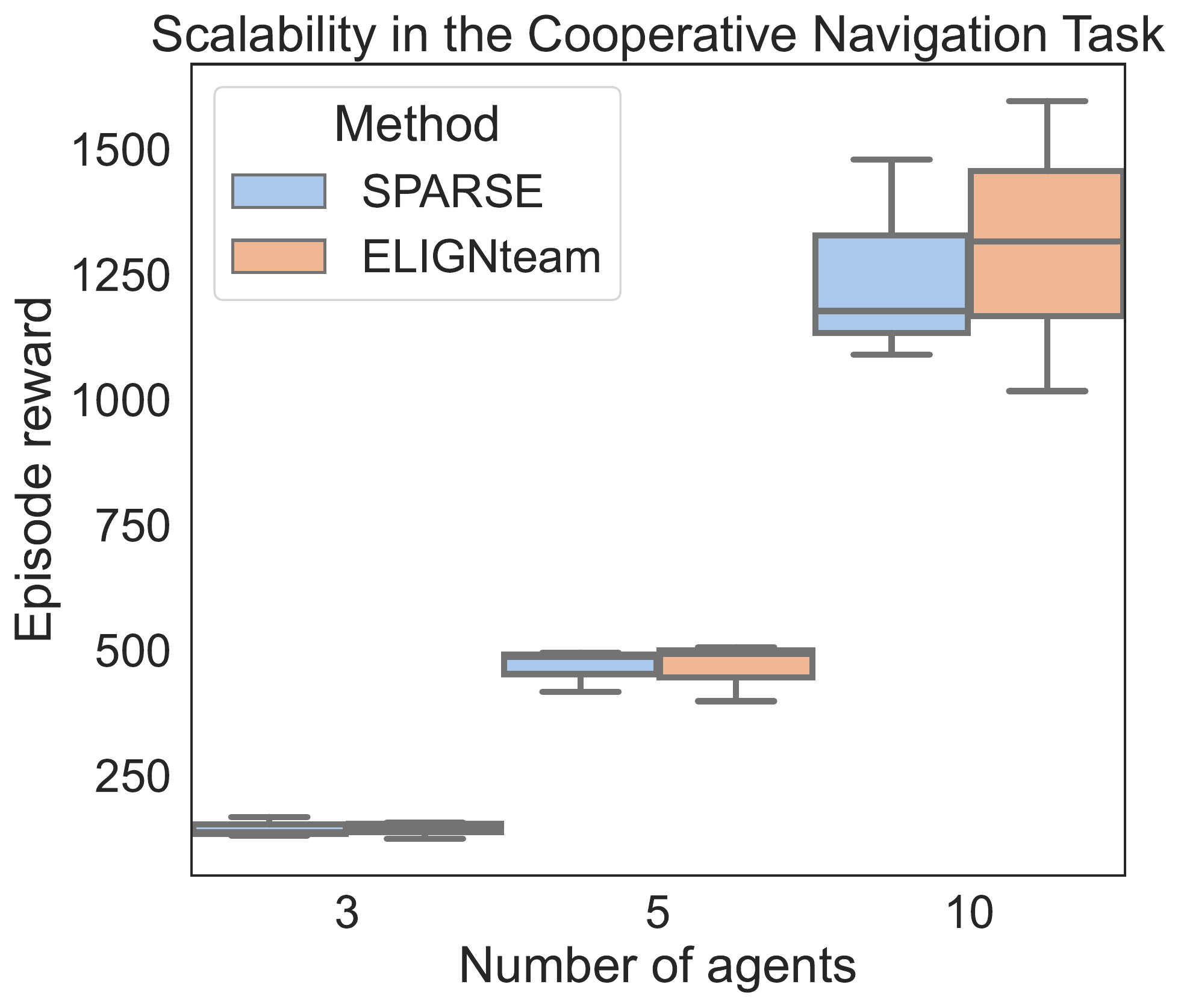}
    \caption{$\textrm{\textsc{elign}}_\textrm{team}$ achieves consistent gains compared against \textsc{sparse} when the number of agents increases in the \textit{Cooperative navigation} task.}
    \label{fig:scalability_plot}
\end{wrapfigure}

\textbf{When the number of agents increases, ELIGN scales well in all multi-agent particle tasks except for \textit{Heterogenous navigation}.}
Table~\ref{tab:3_scaled_dec_reward} shows that our $\textrm{\textsc{elign}}$ intrinsic reward still largely achieves the best performance when more agents are added to cooperative and competitive tasks. The only exception is the \textit{Heterogenous navigation} task, where both $\textrm{\textsc{elign}}_\textrm{self}$ and $\textrm{\textsc{elign}}_\textrm{team}$ outperform \textsc{sparse} but not $\textrm{\textsc{curio}}_\textrm{self/team}$. We hypothesize that it is more difficult for agents to predict their neighbors' future states accurately when there are more agents with different sizes and speeds, and errors in dynamics prediction could lead to misleading alignment signals. Further, we see a consistent increase in $\textrm{\textsc{elign}}_\textrm{team}$'s performance compared against \textsc{sparse} even when the number of agents scales to ten in \textit{Cooperative navigation} (Figure~\ref{fig:scalability_plot}).

\subsection{Results with full observability and centralized ELIGN}
We further test the utility of decentralized ELIGN in fully observable environments and centralized ELIGN under partial observability. We find that decentralized expectation alignment helps in fully observable \textit{Cooperative navigation}, \textit{Heterogenous navigation}, and \textit{Predator-prey}, tasks where expectation alignment has been observed in nature. Similarly, centralized ELIGN also improves agents' performance compared against \textsc{sparse} and $\textrm{\textsc{curio}}_\textrm{self/team}$ rewards in the same tasks with partial observability. These results can be found in Tables~ 3 and 4
in the Appendix. As full observability and centralized training are our main focus, we leave it to future work to investigate why expectation alignment benefits these tasks but not others. 

\begin{wrapfigure}[10]{R}{0.6\linewidth}
    \centering
    \vspace{-2\intextsep}
 \includegraphics[width=\linewidth]{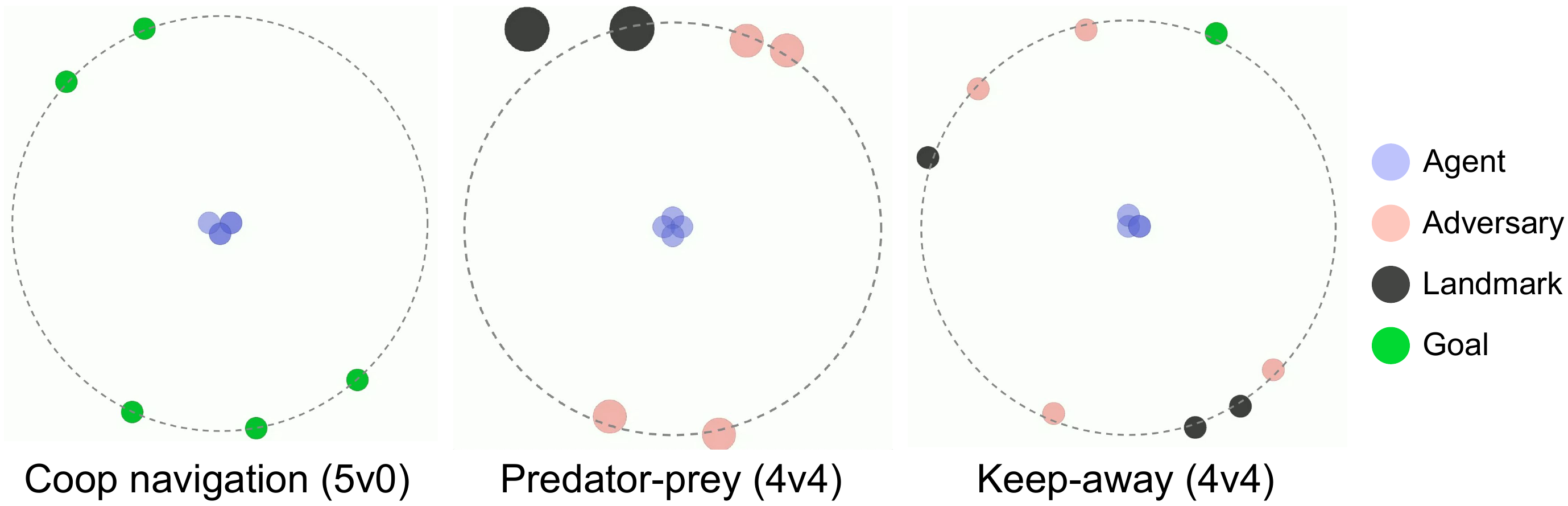}
    \caption{We visualize the symmetry-breaking setups in three example tasks. More details can be found in the Appendix.
    }
    \label{fig:symmetry_setups}
\end{wrapfigure}

\subsection{Investigating how the ELIGN reward helps}
We further investigate how expectation alignment improves coordination through three evaluation setups.

\begin{figure*}
    \centering
    \includegraphics[width=\linewidth]{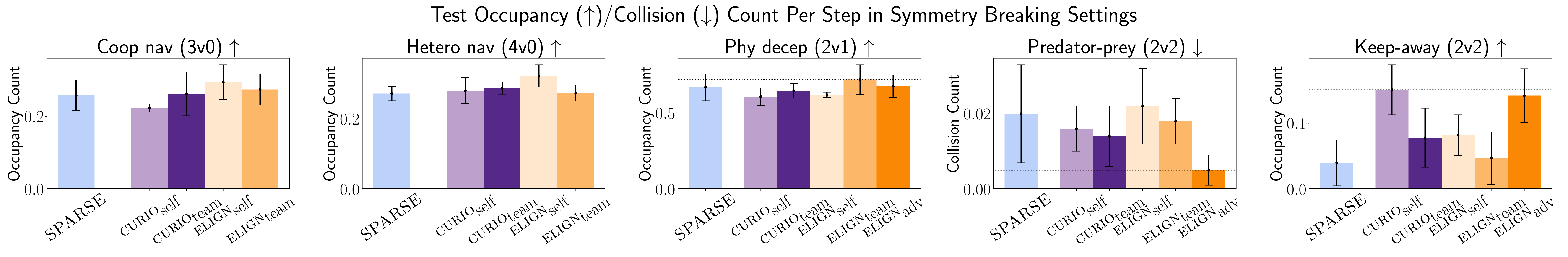}
    \vspace{-1em}
    \caption{We plot the test occupancy/collision count per step of decentralized algorithms in symmetry-breaking settings under partial observability. $\textrm{\textsc{elign}}_\textrm{self}$ performs the best in both cooperative tasks. $\textrm{\textsc{elign}}_\textrm{team}$ and $\textrm{\textsc{elign}}_\textrm{adv}$ are the best strategies in \textit{Physical deception} and \textit{Predator-prey}.
    }
    \label{fig:symmetry_po_occ}
\end{figure*}

\noindent\textbf{ELIGN helps agents divide sub-tasks.} 
A core challenge in multi-agent collaboration is efficient task division~\citep{wang2020cooks}. Here, we test whether expectation alignment improves sub-task allocation. We initialize agents in states without an optimal sub-task allocation, necessitating symmetry-breaking~\citep{hu2020other}. 
Figure~\ref{fig:symmetry_setups} illustrates the symmetry-breaking setups: in cooperative navigation, when agents are initialized equidistant to all the goal locations, there isn't an optimal allocation of agents to goals. 
We find that $\textrm{\textsc{elign}}_\textrm{self}$ achieves the best performance in both cooperative tasks, while $\textrm{\textsc{elign}}_\textrm{team}$ and $\textrm{\textsc{elign}}_\textrm{adv}$ are the best strategies in \textit{Physical deception} and \textit{Predator-prey} respectively (Figure~\ref{fig:symmetry_po_occ}). Upon a qualitative evaluation of \textit{Cooperative navigation}, we observe that agents with expectation alignment are able to predict which goals will be covered by their collaborators and move towards their allocated one. Without expectation alignment, agents move towards the same goal.

\begin{figure*}
    \centering
    \includegraphics[width=\linewidth]{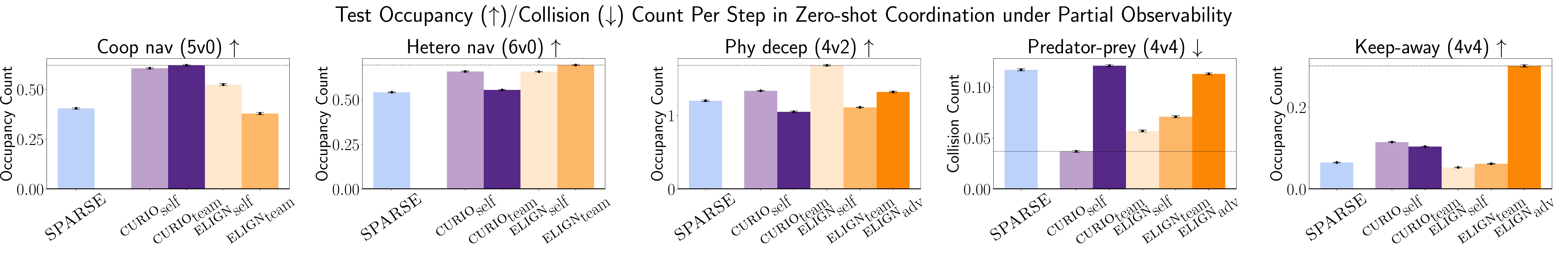}
    \vspace{-2em}
    \caption{
    We plot the test occupancy/collision count per step of decentralized algorithms in zero-shot evaluation with new partners. 
    \review{We find that one of $\textrm{\textsc{elign}}_\textrm{self,team,adv}$ achieves the best performance in \textit{Heterogenous navigation}, \textit{Physical deception} and \textit{Keep away}.}
    }
    \label{fig:zero_shot_po_occ}
\end{figure*}

\noindent\textbf{ELIGN helps agents generalize to new partners.}
Another core challenge in multi-agent collaboration is zero-shot coordination, where agents are tested to collaborate with new partners they haven't been trained with. 
We study whether expectation alignment enables better zero-shot coordination. New partners are sampled from other training runs with different seeds and the team is evaluated using the same metrics as before. 
\review{We observe that \textsc{elign} strategies enable better performance than \textsc{sparse} on average, and one of $\textrm{\textsc{elign}}_\textrm{self,team,adv}$ performs the best in \textit{Heterogenous navigation}, \textit{Physical deception} and \textit{Keep away}. (Figure~\ref{fig:zero_shot_po_occ}). }
These results suggest that ELIGN results in better zero-shot coordination with new partners sampled from separate training runs.  

\noindent\textbf{Accuracy of the dynamics model affects ELIGN.} 
\begin{figure}
    \centering
    \includegraphics[width=0.57\linewidth]{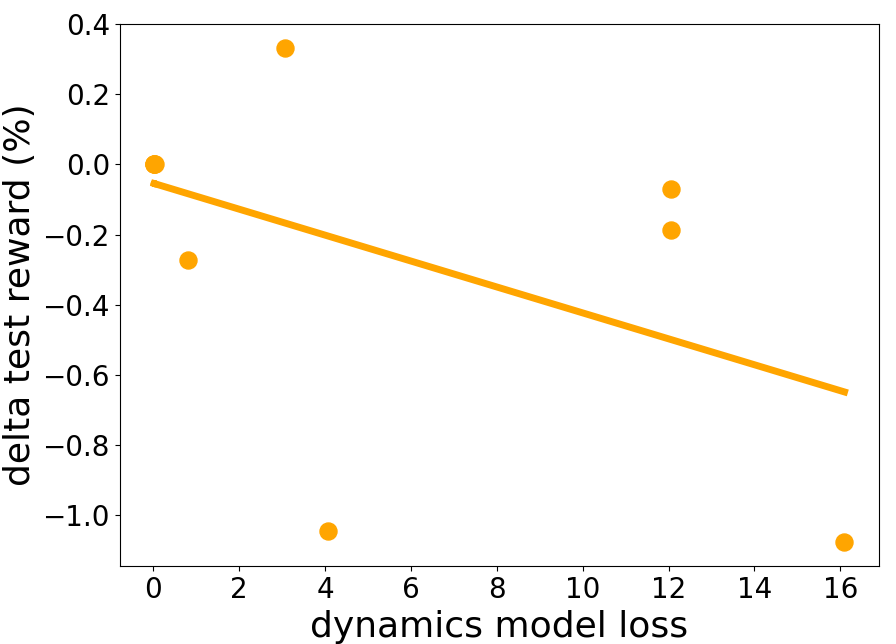}
    \vspace{-1em}
    \caption{Test performance decreases with dynamics model loss ($R^2 = 0.242$), implying that ELIGN requires an accurate dynamics model.}
    \label{fig:dynamics_model_plot}
\end{figure}
We investigate the accuracy of the dynamics model in calculating useful intrinsic rewards. 
Since ELIGN uses a dynamics model to calculate rewards, we test whether an inaccurate model misleads agents towards unaligned behaviors. We train agents on noisy dynamics models by adding Gaussian noise $\epsilon \sim \mathcal{N}(0, \sigma)$ to the predictions made by the dynamics model. We run experiments with multiple $\sigma$ values to study how performance changes as the dynamics model becomes more noisy: $\sigma \in [0.5,1.0, 2.0]$. Our experiments cover one cooperative and one competitive task. Figure~\ref{fig:dynamics_model_plot} plots the \textit{final} dynamics loss against the reward change from the noiseless run. 
As the dynamics model degrades, we observe that the task performance also drops. This study identifies the importance of an accurate dynamics model, suggesting that expectation alignment should used in environments where an accurate dynamics model can be learned.


\section{Discussion}
\label{sec:discussion}
\noindent\textbf{Limitations and future work.} While curiosity has proven useful for exploration in single-agent tasks, we find that expectation alignment—which mathematically encourages agents to be more predictable instead of finding novelty—outperforms curiosity in numerous multi-agent tasks. We hypothesize that our results arise because today’s multi-agent task state space requires significantly less exploration than those used for single-agent (e.g. Atari games). Our findings are limited to the multi-agent particle and Google Research football environments, which have a smaller action space than most ecologically valid scenarios. 

Language, motion, and human gesture are all combinatorially vast; in such action spaces, expectation alignment might develop social dynamics that hinder non-optimal multi-agent behaviors. Similarly, photorealistic environments have a larger state space, where teams perform common household activities (\eg cooking, cleaning, etc.) or drive together in crowded cities~\citep{srivastava2021behavior}. Future work should develop new multi-agent environments that demand exploration complexity and where both curiosity and expectation alignment would be necessary for collaboration. For example, in a search and rescue task where a single agent is unable to carry the injured, curiosity would encourage ``search’’ while ELIGN would speed up ``rescue’’. In the end, we envision that both these forms of rewards would be necessary for successful collaboration. However, choosing when to encourage curiosity versus expectation alignment is an open research problem.

Additionally, enabling stable multi-agent training without centralized training could open up future opportunities for legible~\citep{dragan2013legibility} agents in human environments.  Agents with interpretable actions can induce more faithful human mental models, improving human-AI interaction; however, predictability does not imply legibility. Future work could explore the role of legibility in designing intrinsic rewards.

 \review{Future work should also explore the use of expectation alignment in massive collaboration settings with hundreds of agents. Drawing on Zoology research~\cite{couzin2007collective} expectation alignment should scale in such settings if agents align their behaviors only to their nearest neighbors and not to the entire cohort.} 



\noindent\textbf{Conclusion.}
Inspired by the self-organizing principle in Zoology, we introduce ELIGN, \ie expectation alignment, a simple, task-agnostic, and self-supervised intrinsic reward for multi-agent systems. ELIGN rewards agents when they act predictably to their teammates and unpredictably to their adversaries. ELIGN improves multi-agent performance across six cooperative and competitive tasks in the multi-agent particle and Google Research football environments, especially for decentralized training under partial observability. It also scales well, helps agents break symmetries, and generalize to new partners. 

\bibliographystyle{icml2022}
\bibliography{references}

\section*{Checklist}

\begin{enumerate}

\item For all authors...
\begin{enumerate}
  \item Do the main claims made in the abstract and introduction accurately reflect the paper's contributions and scope?
    \answerYes{See Section~\ref{sec:experiments} for the results mentioned in the abstract.}
  \item Did you describe the limitations of your work?
    \answerYes{See Section~\ref{sec:discussion} for a discussion on the limitations.}
  \item Did you discuss any potential negative societal impacts of your work?
    \answerYes{See the Appendix }
  \item Have you read the ethics review guidelines and ensured that your paper conforms to them?
    \answerYes{We have read the ethics review guidelines.}
\end{enumerate}

\item If you are including theoretical results...
\begin{enumerate}
  \item Did you state the full set of assumptions of all theoretical results?
    \answerNA{}
        \item Did you include complete proofs of all theoretical results?
    \answerNA{}
\end{enumerate}

\item If you ran experiments...
\begin{enumerate}
  \item Did you include the code, data, and instructions needed to reproduce the main experimental results (either in the supplemental material or as a URL)?
    \answerYes{See the Appendix }
  \item Did you specify all the training details (e.g., data splits, hyperparameters, how they were chosen)?
    \answerYes{See the Experiments~\ref{sec:experiments} and Appendix  sections.}
        \item Did you report error bars (e.g., with respect to the random seed after running experiments multiple times)?
    \answerYes{See the Experiments~\ref{sec:experiments} section.}
        \item Did you include the total amount of compute and the type of resources used (e.g., type of GPUs, internal cluster, or cloud provider)?
  \answerYes{See the Experiments~\ref{sec:experiments} and Appendix  sections.}
\end{enumerate}

\item If you are using existing assets (e.g., code, data, models) or curating/releasing new assets...
\begin{enumerate}
  \item If your work uses existing assets, did you cite the creators?
    \answerYes{See the Alignment~\ref{sec:model} and Experiments~\ref{sec:experiments} sections.}
  \item Did you mention the license of the assets?
    \answerYes{See the Appendix }
  \item Did you include any new assets either in the supplemental material or as a URL?
    \answerYes{See the Appendix }
  \item Did you discuss whether and how consent was obtained from people whose data you're using/curating?
    \answerNA{We did not collect any data.}
  \item Did you discuss whether the data you are using/curating contains personally identifiable information or offensive content?
    \answerNA{We did not collect any data.}
\end{enumerate}

\item If you used crowdsourcing or conducted research with human subjects...
\begin{enumerate}
  \item Did you include the full text of instructions given to participants and screenshots, if applicable?
    \answerNA{We did not crowdsource any data.}
  \item Did you describe any potential participant risks, with links to Institutional Review Board (IRB) approvals, if applicable?
    \answerNA{There were no human subjects experiments in our work.}
  \item Did you include the estimated hourly wage paid to participants and the total amount spent on participant compensation?
    \answerNA{No human subjects were hired.}
\end{enumerate}

\end{enumerate}



\appendix

\appendix
\section{Appendix}
\label{sec:appendix}
\subsection{Code}
We upload our code for training and evaluating agents with and without expectation alignment in both the multi-agent particle and Google Research football environments here: \url{https://github.com/StanfordVL/alignment}.



\subsection{Symmetry-breaking initializations}
We create a symmetry-breaking version of each task for evaluation by initializing the environment in the following ways:

\textit{Cooperative Navigation and Heterogenous Navigation:} All agents are initialized at the origin (i.e. center of the world), and target landmarks are placed randomly on a circle perimeter with the maximum radius (i.e. world radius - the greatest landmark size) so that each agent is equidistant from each target landmark. 

\textit{Physical Deception:} Both agents and adversaries start at the origin. All the landmarks, including the goal, are randomly initialized on a circle perimeter.  

\textit{Predator-prey:} The collaborative agents are initialized at the center while the adversaries are placed randomly on a circle perimeter. All the landmarks are randomly initialized in the world.

\textit{Keep-away:} All the cooperative agents are placed at the origin. Adversaries and landmarks, including the goal, are randomly initialized on a circle perimeter. In this task setup, we do not initialize the adversaries at the center because they are awarded for colliding with the cooperative agents.


\review{\subsection{Learning curves}

\begin{figure*}
    \centering
    \includegraphics[width=\linewidth]{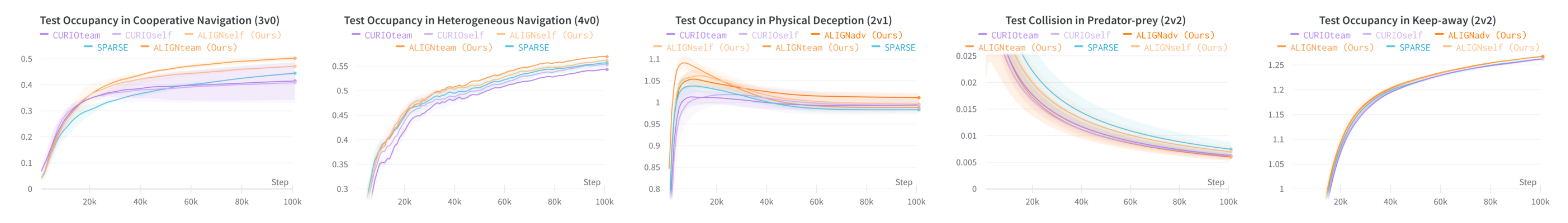}
    \caption{
Learning curves of test occupancy/collision in all five tasks in the multi-agent particle environment. On average, it takes the best $\textrm{\textsc{elign}}$ variant 65 epochs to reach the maximum score of the best $\textrm{\textsc{curio}}$ method at 100 epochs, which means $\textrm{\textsc{elign}}$ requires on average 35\% fewer samples than $\textrm{\textsc{curio}}$.
    }
    \label{fig:learning_curves_occ}
\end{figure*}
\begin{figure*}
    \centering
    \includegraphics[width=1.0\linewidth]{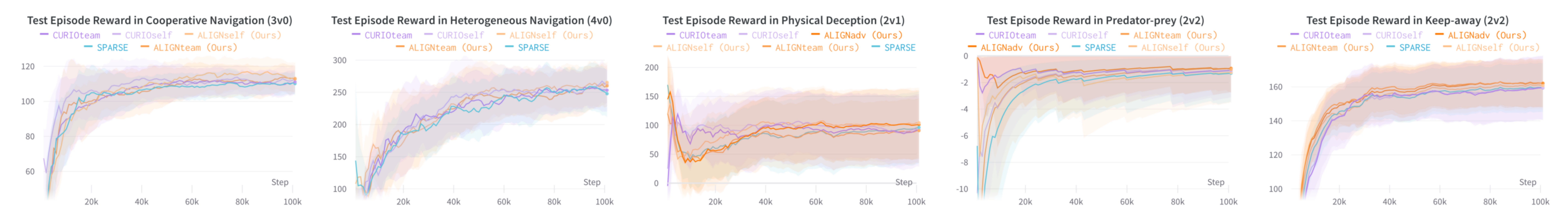}
    \caption{
Learning curves of test episode rewards in all five tasks in the multi-agent particle environment.
    }
    \label{fig:learning_curves_rew}
\end{figure*}
Compared to the best baseline’s highest performance at the 100th epoch, we find that it only takes the best ELIGN variant 31 (Coop nav), 74 (Hetero nav), 30 (Physical dec), 97 (Predator-prey) and 92 (Keep-away) epochs respectively in the five multi-agent particle tasks (\ref{fig:learning_curves_occ}). This means that on average the best ELIGN variant requires $35\pm 32$ fewer training steps to reach the same performance as curiosity or spare methods. 
}

\subsection{Assets and licenses}
We use four assets in total, two of which are existing multi-agent reinforcement learning environments, and the other two are libraries for training reinforcement learning algorithms. 

We conduct our evaluation on the multi-agent particle \citep{lowe2017multi} and Google Research football \citep{kurach2019gfootball} environments, which are under the MIT license and Apache-2.0 license respectively.

We adapt the tianshou \citep{weng2021tianshou} and rllib (part of the ray package) \citep{liang2017rllib} libraries to our experiments, and they are under the MIT license and Apache-2.0 license respectively.

\subsection{Societal impacts}
While developing new intrinsic rewards to improve decentralized multi-agent training can help develop and deploy agents in a variety of applications, we foresee no immediate societal consequences of this work.
However, our experiments thus far have not studied the possible degradation of behaviors when agents align to malicious teammates.
We have also not tested how emergent properties promote better or worse human collaborators.

\subsection{Additional tables}
We include 20 tables of additional results that quantify the agents' performance under fully observable environments, with centralized training, and beyond extrinsic reward. 

Table~\ref{tab:4_dec_reward_fo} reports the test episode rewards of decentralized methods under the fully observable multi-agent particle environment.

Table~\ref{tab:5_cen_reward} reports the test episode rewards of centralized methods under partial observability.

Table~\ref{tab:4_dec_occupancy} and ~\ref{tab:5_dec_distance} report two sets of metrics of \textit{decentralized} methods trained with different intrinsic rewards in both partially and fully observable settings. Table~\ref{tab:4_dec_occupancy} reports the average number of agent-target occupancies per step (or, we can understand it as: on average, the total number of goals occupied by the agents at any given timestep throughout an episode) and agent-adversary collisions in \textit{Predator-prey}. Higher scores are better for the occupancy metric, and lower scores are better for collision. Table ~\ref{tab:5_dec_distance} reports the average minimum agent-to-target distance and agent-to-adversary distance. Agent-to-target distances measure the closest distance an agent achieves to the target location; lower scores are better on this metric. Agent-to-adversary distances measure the closest distance an adversary gets to a good agent; higher scores are better on this metric. Note that these distance-based metrics are not included in the reward functions, and should mainly be used to make comparisons in the case where primary metrics (\ie reward and occupancy/collision count) have the same values.

Table~\ref{tab:6_scaled_dec_occupancy} and ~\ref{tab:7_scaled_dec_distance} report the same  metrics as ~\ref{tab:4_dec_occupancy} and ~\ref{tab:5_dec_distance} respectively, but in \textit{scaled} environments with more agents.

Tables~\ref{tab:14_amb_dec_reward}, ~\ref{tab:15_amb_dec_occupancy} and ~\ref{tab:16_dec_distance} report the test episode reward and additional metrics of \textit{decentralized} algorithms in the \textit{symmetry-breaking} experiments conducted under ``Investigating how alignment reward helps''. Table ~\ref{tab:17_amb_scaled_dec_reward}, ~\ref{tab:18_amb_scaled_dec_occupancy} and ~\ref{tab:19_amb_scaled_dec_distance} report the same set of metrics but from experiments conducted in \textit{scaled} and \textit{symmetry-breaking} environments.

Table~\ref{tab:8_central_reward_fo} reports the test mean episode rewards of \textit{centralized} algorithms with different intrinsic rewards under full observability. Table ~\ref{tab:9_central_occupancy} and ~\ref{tab:10_central_distance} show the other two sets of metrics (\ie occupancy/collision count and agent-target/agent-adversary distance) of \textit{centralized} algorithms. Table ~\ref{tab:11_scaled_central_reward}, ~\ref{tab:12_scaled_central_occupancy}, and ~\ref{tab:13_scaled_central_distance} contain the same metrics as ~\ref{tab:8_central_reward_fo}, ~\ref{tab:9_central_occupancy} and ~\ref{tab:10_central_distance} respectively, but in \textit{scaled} environments.


Finally, Tables~\ref{tab:20_zero_shot_scaled_dec_reward} and \ref{tab:21_zero_shot_scaled_dec_distance} report the test episode reward values and secondary distance-based metrics for the zero-shot generalization experiments conducted under ``Investigating how ELIGN reward helps''. These experiments measure how well agents trained on different seeds generalized to new partners trained on other seeds.

\begin{table*}[t]
\centering 
\caption{\label{tab:4_dec_reward_fo}We report the mean test episode extrinsic rewards and standard errors of $\textit{decentralized}$ methods with different intrinsic rewards in fully observable environments. 
}
\resizebox{\textwidth}{!}{
\begin{tabular}[width=\linewidth]{llrp{0.001pt}rp{0.001pt} rp{0.001pt}rp{0.001pt}r}
\multicolumn{2}{l}{}                & \multicolumn{3}{c}{\textbf{Cooperative}}                              & & \multicolumn{5}{c}{\textbf{Competitive}}                                                                            \\
\cmidrule{3-5}\cmidrule{7-11}
\multicolumn{2}{l}{Task (Agt \# vs. Adv \#) }                & \multicolumn{1}{c}{\textbf{Coop nav. (3v0)}} && \multicolumn{1}{c}{\textbf{Hetero nav. (4v0)}} && \multicolumn{1}{c}{\textbf{Phy decep. (2v1)}}  && \multicolumn{1}{c}{\textbf{Pred-prey (2v2)}} && \multicolumn{1}{c}{\textbf{Keep-away (2v2)}} \\
\cmidrule{1-2}\cmidrule{3-3}\cmidrule{5-5}\cmidrule{7-7}\cmidrule{9-9}\cmidrule{11-11}
\multirow{3}{*}{\textbf{Full observability}}    & $\textsc{sparse}^1$      & $154.00\pm10.51$           && $274.75\pm19.74$            && $82.97\pm 12.23$                && $-10.48\pm 4.20$              && $\mathbf{4.95}\pm \mathbf{2.96}$            
                         \\
                         & $\textrm{\textsc{curio}}_\textrm{self}$   & $154.71\pm 8.00$            && $268.85\pm 15.61$            && $\mathbf{100.66}\pm \mathbf{15.14}$                && $-8.74\pm4.62$              && $-2.00\pm 1.24$            \\
                         & $\textrm{\textsc{elign}}_\textrm{self}$ &  $\mathbf{161.70}\pm \mathbf{4.52}$            && $\mathbf{280.16}\pm\mathbf{17.12}$           && $87.50 \pm 15.40$                && $\mathbf{-5.60}\pm \mathbf{2.60}$              && $0.40\pm 1.92$ \\
                               \multicolumn{11}{l}{$^1$~\cite{lowe2017multi}} 
\end{tabular}
}
\end{table*}

\begin{table*}[t]
\centering 
\caption{\label{tab:5_cen_reward}We report the mean test episode extrinsic rewards and standard errors of $\textit{centralized}$ methods with different intrinsic rewards under partial and full observability. 
}
\resizebox{\textwidth}{!}{
\begin{tabular}{llrp{0.001pt}rp{0.001pt} rp{0.001pt}rp{0.001pt}r}
\multicolumn{2}{l}{}                & \multicolumn{3}{c}{\textbf{Cooperative}}                              & & \multicolumn{5}{c}{\textbf{Competitive}}                                                                            \\
\cmidrule{3-5}\cmidrule{7-11}
\multicolumn{2}{l}{Task (Agt \# vs. Adv \#) }                & \multicolumn{1}{c}{\textbf{Coop nav. (3v0)}} && \multicolumn{1}{c}{\textbf{Hetero nav. (4v0)}} && \multicolumn{1}{c}{\textbf{Phy decep. (2v1)}}  && \multicolumn{1}{c}{\textbf{Pred-prey (2v2)}} && \multicolumn{1}{c}{\textbf{Keep-away (2v2)}} \\
\cmidrule{1-2}\cmidrule{3-3}\cmidrule{5-5}\cmidrule{7-7}\cmidrule{9-9}\cmidrule{11-11}
\multirow{6}{*}{\textbf{Partial observability}} & \textsc{sparse}$^1$                & $113.25\pm8.10 $ && $ 178.62\pm9.62 $ && $ \mathbf{117.45}\pm\mathbf{10.63} $ && $ -1.96\pm1.45 $ && $ \mathbf{35.79}\pm\mathbf{14.93} $\\
                         &  $\textrm{\textsc{curio}}_\textrm{self}^2$        & $128.77\pm7.70 $ && $ 190.30\pm7.73 $ && $ 111.08\pm10.09 $ && $ -1.63\pm1.27 $ && $ 13.94\pm12.56 $           \\
                         & $\textrm{\textsc{curio}}_\textrm{team}^3$           & $114.13\pm11.84 $ && $ 189.80\pm11.81 $ && $ 114.32\pm5.46 $ && $ -3.04\pm1.09 $ && $ 6.01\pm3.36$\\
                         & $\textrm{\textsc{elign}}_\textrm{self}$        & $\mathbf{137.14}\pm\mathbf{3.63} $ && $ 169.58\pm14.99 $ && $ 93.27\pm3.70 $ && $ -0.41\pm0.28 $ && $ 22.77\pm9.91$            \\
                         & $\textrm{\textsc{elign}}_\textrm{team}$  & $119.10\pm10.89 $ && $ \mathbf{210.81}\pm\mathbf{9.70} $ && $ 96.49\pm6.46 $ && $ -0.92\pm0.72 $ && $ 24.94\pm12.58 $                     \\
                         & $\textrm{\textsc{elign}}_\textrm{adv}$      & ---          &&  ---                && $ 102.37\pm6.98 $ && $ \mathbf{-0.13}\pm\mathbf{0.03} $ && $ 8.70\pm4.44 $\\
                        \multicolumn{11}{l}{$^1$~\cite{lowe2017multi},$^2$~\cite{stadie2015incentivizing},$^3$~\cite{iqbal2020coordinated}}  
\end{tabular}
}
\end{table*}

\begin{table*}[t]
\centering 
\caption{\label{tab:4_dec_occupancy} The average test occupancy/collision count per step and standard errors of $\textit{decentralized}$ methods with different intrinsic rewards under partial and full observability. Higher scores are better for the occupancy metric ($\uparrow$), and lower scores are better for the collision metric ($\downarrow$). 
}
\resizebox{\textwidth}{!}{
\begin{tabular}{llrp{0.001pt}rp{0.001pt} rp{0.001pt}rp{0.001pt}r}
\multicolumn{2}{l}{}                & \multicolumn{3}{c}{\textbf{Cooperative}}                              & & \multicolumn{5}{c}{\textbf{Competitive}}                                                                            \\
\cmidrule{3-5}\cmidrule{7-11}
\multicolumn{2}{l}{Task (Agt \# vs. Adv \#) }                & \multicolumn{1}{c}{\textbf{Coop nav. (3v0)} $\uparrow$} && \multicolumn{1}{c}{\textbf{Hetero nav. (4v0)} $\uparrow$} && \multicolumn{1}{c}{\textbf{Phy decep. (2v1)} $\uparrow$}  && \multicolumn{1}{c}{\textbf{Pred-prey (2v2)} $\downarrow$} && \multicolumn{1}{c}{\textbf{Keep-away (2v2)} $\uparrow$} \\
\cmidrule{1-2}\cmidrule{3-3}\cmidrule{5-5}\cmidrule{7-7}\cmidrule{9-9}\cmidrule{11-11}
\multirow{6}{*}{\textbf{Partial observability}}  &$\textsc{sparse}$&$0.46\pm0.05$ && $0.57\pm0.01$ && $0.98\pm0.07$ && $0.02\pm0.01$ && $0.07\pm0.02$\\
&$\textrm{\textsc{curio}}_\textrm{self}$&$0.43\pm0.03$ && $0.60\pm0.01$ && $0.99\pm0.03$ && $0.02\pm0.01$ && $0.14\pm0.02$\\
&$\textrm{\textsc{curio}}_\textrm{team}$&$0.42\pm0.05$ && $0.59\pm0.01$ && $0.95\pm0.01$ && $0.02\pm0.01$ && $0.10\pm0.03$\\
&$\textrm{\textsc{elign}}_\textrm{self}$&$0.52\pm0.03$ && $0.61\pm0.01$ && $0.95\pm0.02$ && $0.03\pm0.01$ && $0.10\pm0.02$\\
&$\textrm{\textsc{elign}}_\textrm{team}$&$0.44\pm0.04$ && $0.58\pm0.02$ && $0.99\pm0.07$ && $0.03\pm0.01$ && $0.07\pm0.02$\\
&$\textrm{\textsc{elign}}_\textrm{adv}$&--- && --- && $1.00\pm0.06$ && $0.01\pm0.01$ && $0.15\pm0.03$\\

                         \hline
\multirow{3}{*}{\textbf{Full observability}}   
&$\textsc{sparse}$&$0.46\pm0.11$ && $0.57\pm0.01$ && $0.88\pm0.09$ && $0.03\pm0.01$ && $0.06\pm0.02$\\
&$\textrm{\textsc{curio}}_\textrm{self}$&$0.50\pm0.07$ && $0.59\pm0.02$ && $1.09\pm0.13$ && $0.03\pm0.01$ && $0.02\pm0.00$\\
&$\textrm{\textsc{elign}}_\textrm{self}$&$0.48\pm0.11$ && $0.58\pm0.02$ && $0.83\pm0.10$ && $0.02\pm0.01$ && $0.04\pm0.01$\\
\end{tabular}
}
\end{table*}

\begin{table*}[t]
\centering 
\caption{\label{tab:5_dec_distance} The average test agent-to-target (agt-target) and agent-to-adversary (agt-adv) distances and standard errors of $\textit{decentralized}$ methods with different intrinsic rewards under partial and full observability. Lower scores are better for agt-target ($\downarrow$), and higher scores are better for agt-adv ($\uparrow$).
}
\resizebox{\textwidth}{!}{
\begin{tabular}{llrp{0.001pt}rp{0.001pt} rp{0.001pt}rp{0.001pt}r}
\multicolumn{2}{l}{}                & \multicolumn{3}{c}{\textbf{Cooperative}}                              & & \multicolumn{5}{c}{\textbf{Competitive}}                                                                            \\
\cmidrule{3-5}\cmidrule{7-11}
\multicolumn{2}{l}{Task (Agt \# vs. Adv \#) }                & \multicolumn{1}{c}{\textbf{Coop nav. (3v0)} $\downarrow$} && \multicolumn{1}{c}{\textbf{Hetero nav. (4v0)} $\downarrow$} && \multicolumn{1}{c}{\textbf{Phy decep. (2v1)} $\downarrow$}  && \multicolumn{1}{c}{\textbf{Pred-prey (2v2)} $\uparrow$} && \multicolumn{1}{c}{\textbf{Keep-away (2v2)} $\downarrow$} \\
\cmidrule{1-2}\cmidrule{3-3}\cmidrule{5-5}\cmidrule{7-7}\cmidrule{9-9}\cmidrule{11-11}
\multirow{6}{*}{\textbf{Partial observability}}  &$\textsc{sparse}$&$0.30\pm0.02$ && $0.23\pm0.00$ && $0.26\pm0.01$ && $1.45\pm0.11$ && $1.41\pm0.07$\\
&$\textrm{\textsc{curio}}_\textrm{self}$&$0.32\pm0.02$ && $0.25\pm0.01$ && $0.25\pm0.00$ && $1.36\pm0.06$ && $1.14\pm0.09$\\
&$\textrm{\textsc{curio}}_\textrm{team}$&$0.31\pm0.01$ && $0.25\pm0.01$ && $0.26\pm0.00$ && $1.48\pm0.13$ && $1.31\pm0.10$\\
&$\textrm{\textsc{elign}}_\textrm{self}$&$0.33\pm0.03$ && $0.25\pm0.01$ && $0.26\pm0.00$ && $1.39\pm0.12$ && $1.26\pm0.09$\\
&$\textrm{\textsc{elign}}_\textrm{team}$&$0.33\pm0.02$ && $0.23\pm0.01$ && $0.25\pm0.01$ && $1.38\pm0.13$ && $1.38\pm0.09$\\
&$\textrm{\textsc{elign}}_\textrm{adv}$&--- && --- && $0.25\pm0.01$ && $1.54\pm0.08$ && $1.14\pm0.09$\\
                         \hline
\multirow{3}{*}{\textbf{Full observability}}   
&$\textsc{sparse}$&$0.32\pm0.09$ && $0.23\pm0.00$ && $0.26\pm0.01$ && $1.23\pm0.12$ && $1.27\pm0.09$\\
&$\textrm{\textsc{curio}}_\textrm{self}$&$0.28\pm0.04$ && $0.22\pm0.01$ && $0.23\pm0.01$ && $1.37\pm0.15$ && $1.53\pm0.03$\\
&$\textrm{\textsc{elign}}_\textrm{self}$&$0.30\pm0.07$ && $0.23\pm0.01$ && $0.27\pm0.01$ && $1.40\pm0.13$ && $1.41\pm0.10$\\
\end{tabular}
}
\end{table*}

\begin{table*}[t]
\centering 
\caption{\label{tab:6_scaled_dec_occupancy} The average test occupancy/collision count per step and standard errors of $\textit{decentralized}$ methods with different intrinsic rewards in \textit{scaled} environments under partial and full observability. Higher scores are better for the occupancy metric ($\uparrow$), and lower scores are better for the collision metric ($\downarrow$). 
}
\resizebox{\textwidth}{!}{
\begin{tabular}{llrp{0.001pt}rp{0.001pt} rp{0.001pt}rp{0.001pt}r}
\multicolumn{2}{l}{}                & \multicolumn{3}{c}{\textbf{Cooperative}}                              & & \multicolumn{5}{c}{\textbf{Competitive}}                                                                            \\
\cmidrule{3-5}\cmidrule{7-11}
\multicolumn{2}{l}{Task (Agt \# vs. Adv \#) }                & \multicolumn{1}{c}{\textbf{Coop nav. (5v0)} $\uparrow$} && \multicolumn{1}{c}{\textbf{Hetero nav. (6v0)} $\uparrow$} && \multicolumn{1}{c}{\textbf{Phy decep. (4v2)} $\uparrow$}  && \multicolumn{1}{c}{\textbf{Pred-prey (4v4)} $\downarrow$} && \multicolumn{1}{c}{\textbf{Keep-away (4v4)} $\uparrow$} \\
\cmidrule{1-2}\cmidrule{3-3}\cmidrule{5-5}\cmidrule{7-7}\cmidrule{9-9}\cmidrule{11-11}
\multirow{4}{*}{\textbf{Partial observability}}  &$\textsc{sparse}$&$0.50\pm0.04$ && $0.46\pm0.08$ && $1.20\pm0.10$ && $0.11\pm0.02$ && $0.08\pm0.02$\\
&$\textrm{\textsc{curio}}_\textrm{self}$&$0.48\pm0.03$ && $0.63\pm0.01$ && $1.20\pm0.08$ && $0.07\pm0.02$ && $0.15\pm0.06$\\
&$\textrm{\textsc{curio}}_\textrm{team}$&$0.53\pm0.03$ && $0.60\pm0.02$ && $1.20\pm0.09$ && $0.05\pm0.02$ && $0.06\pm0.01$\\
&$\textrm{\textsc{elign}}_\textrm{self}$&$0.49\pm0.03$ && $0.67\pm0.00$ && $1.30\pm0.23$ && $0.04\pm0.02$ && $0.14\pm0.08$\\
&$\textrm{\textsc{elign}}_\textrm{team}$&$0.56\pm0.04$ && $0.56\pm0.00$ && $1.21\pm0.09$ && $0.08\pm0.02$ && $0.10\pm0.02$\\
&$\textrm{\textsc{elign}}_\textrm{adv}$&--- && --- && $1.23\pm0.10$ && $0.08\pm0.02$ && $0.16\pm0.07$\\

                         \hline
\multirow{2}{*}{\textbf{Full observability}}   
&$\textsc{sparse}$&$0.52\pm0.11$ && $0.46\pm0.08$ && $0.99\pm0.09$ && $0.21\pm0.01$ && $0.03\pm0.00$\\
&$\textrm{\textsc{curio}}_\textrm{self}$&$0.39\pm0.13$ && $0.56\pm0.01$ && $0.86\pm0.04$ && $0.16\pm0.03$ && $0.04\pm0.00$\\
&$\textrm{\textsc{elign}}_\textrm{self}$&$0.55\pm0.11$ && $0.56\pm0.00$ && $1.04\pm0.07$ && $0.15\pm0.03$ && $0.06\pm0.02$\\
\end{tabular}
}
\end{table*}

\begin{table*}[t]
\centering 
\caption{\label{tab:7_scaled_dec_distance} The average test agent-to-target (agt-target) and agent-to-adversary (agt-adv) distances and standard errors of $\textit{decentralized}$ methods with different intrinsic rewards in \textit{scaled} environments under partial and full observability. Lower scores are better for agt-target ($\downarrow$), and higher scores are better for agt-adv ($\uparrow$).
}
\resizebox{\textwidth}{!}{
\begin{tabular}{llrp{0.001pt}rp{0.001pt} rp{0.001pt}rp{0.001pt}r}
\multicolumn{2}{l}{}                & \multicolumn{3}{c}{\textbf{Cooperative}}                              & & \multicolumn{5}{c}{\textbf{Competitive}}                                                                            \\
\cmidrule{3-5}\cmidrule{7-11}
\multicolumn{2}{l}{Task (Agt \# vs. Adv \#) }                & \multicolumn{1}{c}{\textbf{Coop nav. (5v0)} $\downarrow$} && \multicolumn{1}{c}{\textbf{Hetero nav. (6v0)} $\downarrow$} && \multicolumn{1}{c}{\textbf{Phy decep. (4v2)} $\downarrow$}  && \multicolumn{1}{c}{\textbf{Pred-prey (4v4)} $\uparrow$} && \multicolumn{1}{c}{\textbf{Keep-away (4v4)} $\downarrow$} \\
\cmidrule{1-2}\cmidrule{3-3}\cmidrule{5-5}\cmidrule{7-7}\cmidrule{9-9}\cmidrule{11-11}
\multirow{4}{*}{\textbf{Partial observability}}  &$\textsc{sparse}$&$0.22\pm0.01$ && $0.27\pm0.05$ && $0.23\pm0.02$ && $2.03\pm0.15$ && $2.97\pm0.17$\\
&$\textrm{\textsc{curio}}_\textrm{self}$&$0.30\pm0.02$ && $0.21\pm0.01$ && $0.24\pm0.01$ && $2.18\pm0.13$ && $2.70\pm0.25$\\
&$\textrm{\textsc{curio}}_\textrm{team}$&$0.23\pm0.02$ && $0.22\pm0.01$ && $0.23\pm0.02$ && $2.29\pm0.12$ && $3.14\pm0.08$\\
&$\textrm{\textsc{elign}}_\textrm{self}$&$0.29\pm0.03$ && $0.19\pm0.00$ && $0.24\pm0.02$ && $2.39\pm0.11$ && $2.97\pm0.30$\\
&$\textrm{\textsc{elign}}_\textrm{team}$&$0.23\pm0.04$ && $0.21\pm0.00$ && $0.23\pm0.01$ && $2.16\pm0.12$ && $2.88\pm0.19$\\
&$\textrm{\textsc{elign}}_\textrm{adv}$&--- && --- && $0.22\pm0.01$ && $2.12\pm0.16$ && $2.66\pm0.23$\\
                         \hline
\multirow{2}{*}{\textbf{Full observability}}   
&$\textsc{sparse}$&$0.23\pm0.06$ && $0.27\pm0.05$ && $0.21\pm0.02$ && $1.64\pm0.02$ && $3.28\pm0.02$\\
&$\textrm{\textsc{curio}}_\textrm{self}$&$0.33\pm0.09$ && $0.21\pm0.01$ && $0.22\pm0.01$ && $1.81\pm0.12$ && $3.24\pm0.08$\\
&$\textrm{\textsc{elign}}_\textrm{self}$&$0.20\pm0.04$ && $0.21\pm0.00$ && $0.21\pm0.01$ && $1.82\pm0.10$ && $2.97\pm0.17$\\
\end{tabular}
}
\end{table*}

\begin{table*}[t]
\centering 
\caption{\label{tab:14_amb_dec_reward}We report the mean test episode extrinsic rewards and standard errors of $\textit{decentralized}$ methods with different intrinsic rewards in \textit{symmetry-breaking} settings under partial and full observability. 
}
\resizebox{\textwidth}{!}{
\begin{tabular}{llrp{0.001pt}rp{0.001pt} rp{0.001pt}rp{0.001pt}r}
\multicolumn{2}{l}{}                & \multicolumn{3}{c}{\textbf{Cooperative}}                              & & \multicolumn{5}{c}{\textbf{Competitive}}                                                                            \\
\cmidrule{3-5}\cmidrule{7-11}
\multicolumn{2}{l}{Task (Agt \# vs. Adv \#) }                & \multicolumn{1}{c}{\textbf{Coop nav. (3v0)}} && \multicolumn{1}{c}{\textbf{Hetero nav. (4v0)}} && \multicolumn{1}{c}{\textbf{Phy decep. (2v1)}}  && \multicolumn{1}{c}{\textbf{Pred-prey (2v2)}} && \multicolumn{1}{c}{\textbf{Keep-away (2v2)}} \\
\cmidrule{1-2}\cmidrule{3-3}\cmidrule{5-5}\cmidrule{7-7}\cmidrule{9-9}\cmidrule{11-11}
\multirow{6}{*}{\textbf{Partial observability}} &$\textsc{sparse}$&$97.45\pm10.49$ && $184.18\pm7.63$ && $59.39\pm21.10$ && $-1.89\pm1.69$ && $3.85\pm4.25$\\
&$\textrm{\textsc{curio}}_\textrm{self}$&$85.23\pm10.88$ && $184.07\pm9.99$ && $54.17\pm27.40$ && $-2.86\pm1.19$ && $19.57\pm4.92$\\
&$\textrm{\textsc{curio}}_\textrm{team}$&$81.50\pm15.78$ && $141.78\pm20.04$ && $41.12\pm13.37$ && $-2.80\pm1.91$ && $10.21\pm6.34$\\
&$\textrm{\textsc{elign}}_\textrm{self}$&$110.29\pm9.67$ && $176.98\pm6.38$ && $98.90\pm17.71$ && $-4.00\pm2.14$ && $9.47\pm3.99$\\
&$\textrm{\textsc{elign}}_\textrm{team}$&$92.41\pm10.70$ && $187.42\pm11.29$ && $74.06\pm21.58$ && $-2.00\pm1.39$ && $3.32\pm3.04$\\
&$\textrm{\textsc{elign}}_\textrm{adv}$&--- && --- && $87.55\pm15.35$ && $-1.40\pm1.25$ && $13.77\pm3.58$\\
                         \hline
\multirow{3}{*}{\textbf{Full observability}}    &$\textsc{sparse}$&$150.42\pm15.18$ && $250.41\pm14.23$ && $69.06\pm14.06$ && $-7.62\pm3.50$ && $3.50\pm4.00$\\
&$\textrm{\textsc{curio}}_\textrm{self}$&$149.48\pm9.42$ && $241.69\pm19.58$ && $52.69\pm17.97$ && $-10.40\pm6.33$ && $-1.10\pm0.59$\\
&$\textrm{\textsc{elign}}_\textrm{self}$&$152.08\pm6.68$ && $275.69\pm7.49$ && $75.79\pm24.54$ && $-4.44\pm2.05$ && $0.96\pm3.14$\\
\end{tabular}
}
\end{table*}

\begin{table*}[t]
\centering 
\caption{\label{tab:15_amb_dec_occupancy} The average test occupancy/collision count per step and standard errors of $\textit{decentralized}$ methods with different intrinsic rewards in \textit{symmetry-breaking} settings  under partial and full observability. Higher scores are better for the occupancy metric ($\uparrow$), and lower scores are better for the collision metric ($\downarrow$). 
}
\resizebox{\textwidth}{!}{
\begin{tabular}{llrp{0.001pt}rp{0.001pt} rp{0.001pt}rp{0.001pt}r}
\multicolumn{2}{l}{}                & \multicolumn{3}{c}{\textbf{Cooperative}}                              & & \multicolumn{5}{c}{\textbf{Competitive}}                                                                            \\
\cmidrule{3-5}\cmidrule{7-11}
\multicolumn{2}{l}{Task (Agt \# vs. Adv \#) }                & \multicolumn{1}{c}{\textbf{Coop nav. (3v0)} $\uparrow$} && \multicolumn{1}{c}{\textbf{Hetero nav. (4v0)} $\uparrow$} && \multicolumn{1}{c}{\textbf{Phy decep. (2v1)} $\uparrow$}  && \multicolumn{1}{c}{\textbf{Pred-prey (2v2)} $\downarrow$} && \multicolumn{1}{c}{\textbf{Keep-away (2v2)} $\uparrow$} \\
\cmidrule{1-2}\cmidrule{3-3}\cmidrule{5-5}\cmidrule{7-7}\cmidrule{9-9}\cmidrule{11-11}
\multirow{6}{*}{\textbf{Partial observability}}  &$\textsc{sparse}$&$0.26\pm0.04$ && $0.27\pm0.02$ && $0.67\pm0.09$ && $0.02\pm0.01$ && $0.04\pm0.03$\\
&$\textrm{\textsc{curio}}_\textrm{self}$&$0.22\pm0.01$ && $0.28\pm0.04$ && $0.61\pm0.06$ && $0.02\pm0.01$ && $0.15\pm0.04$\\
&$\textrm{\textsc{curio}}_\textrm{team}$&$0.26\pm0.06$ && $0.29\pm0.02$ && $0.65\pm0.05$ && $0.01\pm0.01$ && $0.08\pm0.04$\\
&$\textrm{\textsc{elign}}_\textrm{self}$&$0.29\pm0.05$ && $0.32\pm0.03$ && $0.62\pm0.02$ && $0.02\pm0.01$ && $0.08\pm0.03$\\
&$\textrm{\textsc{elign}}_\textrm{team}$&$0.27\pm0.04$ && $0.27\pm0.02$ && $0.72\pm0.10$ && $0.02\pm0.01$ && $0.05\pm0.04$\\
&$\textrm{\textsc{elign}}_\textrm{adv}$&--- && --- && $0.68\pm0.07$ && $0.00\pm0.00$ && $0.14\pm0.04$\\

                         \hline
\multirow{3}{*}{\textbf{Full observability}}   
&$\textsc{sparse}$&$0.45\pm0.12$ && $0.54\pm0.01$ && $0.89\pm0.11$ && $0.03\pm0.01$ && $0.05\pm0.02$\\
&$\textrm{\textsc{curio}}_\textrm{self}$&$0.48\pm0.08$ && $0.54\pm0.01$ && $1.13\pm0.14$ && $0.04\pm0.02$ && $0.00\pm0.00$\\
&$\textrm{\textsc{elign}}_\textrm{self}$&$0.46\pm0.11$ && $0.54\pm0.01$ && $0.86\pm0.12$ && $0.02\pm0.01$ && $0.02\pm0.02$\\
\end{tabular}
}
\end{table*}

\begin{table*}[t]
\centering 
\caption{\label{tab:16_dec_distance} The average test agent-to-target (agt-target) and agent-to-adversary (agt-adv) distances and standard errors of $\textit{decentralized}$ methods with different intrinsic rewards in \textit{symmetry-breaking} settings under partial and full observability. Lower scores are better for agt-target ($\downarrow$), and higher scores are better for agt-adv ($\uparrow$).
}
\resizebox{\textwidth}{!}{
\begin{tabular}{llrp{0.001pt}rp{0.001pt} rp{0.001pt}rp{0.001pt}r}
\multicolumn{2}{l}{}                & \multicolumn{3}{c}{\textbf{Cooperative}}                              & & \multicolumn{5}{c}{\textbf{Competitive}}                                                                            \\
\cmidrule{3-5}\cmidrule{7-11}
\multicolumn{2}{l}{Task (Agt \# vs. Adv \#) }                & \multicolumn{1}{c}{\textbf{Coop nav. (3v0)} $\downarrow$} && \multicolumn{1}{c}{\textbf{Hetero nav. (4v0)} $\downarrow$} && \multicolumn{1}{c}{\textbf{Phy decep. (2v1)} $\downarrow$}  && \multicolumn{1}{c}{\textbf{Pred-prey (2v2)} $\uparrow$} && \multicolumn{1}{c}{\textbf{Keep-away (2v2)} $\downarrow$} \\
\cmidrule{1-2}\cmidrule{3-3}\cmidrule{5-5}\cmidrule{7-7}\cmidrule{9-9}\cmidrule{11-11}
\multirow{6}{*}{\textbf{Partial observability}}  &$\textsc{sparse}$&$0.53\pm0.02$ && $0.57\pm0.02$ && $0.35\pm0.03$ && $1.49\pm0.14$ && $1.57\pm0.15$\\
&$\textrm{\textsc{curio}}_\textrm{self}$&$0.57\pm0.04$ && $0.55\pm0.04$ && $0.37\pm0.02$ && $1.29\pm0.06$ && $1.07\pm0.17$\\
&$\textrm{\textsc{curio}}_\textrm{team}$&$0.53\pm0.03$ && $0.55\pm0.03$ && $0.35\pm0.02$ && $1.49\pm0.13$ && $1.37\pm0.18$\\
&$\textrm{\textsc{elign}}_\textrm{self}$&$0.68\pm0.05$ && $0.52\pm0.03$ && $0.34\pm0.01$ && $1.39\pm0.13$ && $1.26\pm0.18$\\
&$\textrm{\textsc{elign}}_\textrm{team}$&$0.55\pm0.05$ && $0.56\pm0.02$ && $0.31\pm0.02$ && $1.41\pm0.10$ && $1.53\pm0.17$\\
&$\textrm{\textsc{elign}}_\textrm{adv}$&--- && --- && $0.33\pm0.04$ && $1.61\pm0.08$ && $1.08\pm0.15$\\

                         \hline
\multirow{3}{*}{\textbf{Full observability}}   
&$\textsc{sparse}$&$0.45\pm0.12$ && $0.29\pm0.00$ && $0.25\pm0.01$ && $1.28\pm0.15$ && $1.30\pm0.15$\\
&$\textrm{\textsc{curio}}_\textrm{self}$&$0.33\pm0.06$ && $0.30\pm0.01$ && $0.22\pm0.01$ && $1.47\pm0.17$ && $1.71\pm0.05$\\
&$\textrm{\textsc{elign}}_\textrm{self}$&$0.46\pm0.11$ && $0.30\pm0.00$ && $0.25\pm0.02$ && $1.49\pm0.16$ && $1.53\pm0.16$\\
\end{tabular}
}
\end{table*}

\begin{table*}[t]
\centering 
\caption{\label{tab:17_amb_scaled_dec_reward}We report the mean test episode extrinsic rewards and standard errors of $\textit{decentralized}$ methods with different intrinsic rewards in $\textit{scaled}$ and $\textit{symmetry-breaking}$ settings.
}
\resizebox{\textwidth}{!}{
\begin{tabular}[width=\linewidth]{llrp{0.001pt}rp{0.001pt} rp{0.001pt}rp{0.001pt}r}
\multicolumn{2}{l}{}                & \multicolumn{3}{c}{\textbf{Cooperative}}                              & & \multicolumn{5}{c}{\textbf{Competitive}}                                                                            \\
\cmidrule{3-5}\cmidrule{7-11}
\multicolumn{2}{l}{Task (Agt \# vs. Adv \#) }                & \multicolumn{1}{c}{\textbf{Coop nav. (5v0)}} && \multicolumn{1}{c}{\textbf{Hetero nav. (6v0)}} && \multicolumn{1}{c}{\textbf{Phy decep. (4v2)}}  && \multicolumn{1}{c}{\textbf{Pred-prey (4v4)}} && \multicolumn{1}{c}{\textbf{Keep-away (4v4)}} \\
\cmidrule{1-2}\cmidrule{3-3}\cmidrule{5-5}\cmidrule{7-7}\cmidrule{9-9}\cmidrule{11-11}
\multirow{4}{*}{\textbf{Partial observability}} &$\textsc{sparse}$&$328.24\pm24.17$ && $405.08\pm21.53$ && $172.87\pm32.43$ && $-35.40\pm8.63$ && $1.37\pm3.48$\\
&$\textrm{\textsc{curio}}_\textrm{self}$&$295.48\pm20.54$ && $436.17\pm26.30$ && $202.39\pm26.06$ && $-11.19\pm3.65$ && $9.24\pm8.49$\\
&$\textrm{\textsc{curio}}_\textrm{team}$&$316.33\pm14.44$ && $422.71\pm13.24$ && $229.50\pm28.29$ && $-11.56\pm6.37$ && $-1.29\pm1.58$\\
&$\textrm{\textsc{elign}}_\textrm{self}$&$357.40 \pm	19.52$ && $412.39\pm12.63$ && $129.07\pm51.08$ && $-7.34\pm5.12$ && $11.97\pm13.30$\\
&$\textrm{\textsc{elign}}_\textrm{team}$&$354.14\pm19.53$ && $417.94\pm22.29$ && $184.21\pm23.16$ && $-19.37\pm6.44$ && $4.05\pm5.78$\\
&$\textrm{\textsc{elign}}_\textrm{adv}$&--- && --- && $148.69\pm31.79$ && $-23.42\pm8.32$ && $18.71\pm14.78$\\
                               \hline
\multirow{2}{*}{\textbf{Full observability}} &$\textsc{sparse}$&$466.17\pm28.16$ && $471.19\pm16.23$ && $233.61\pm25.44$ && $-39.24\pm6.63$ && $-5.10\pm0.26$\\
&$\textrm{\textsc{curio}}_\textrm{self}$&$509.91\pm14.10$ && $606.07\pm7.55$ && $256.13\pm41.13$ && $-38.66\pm13.38$ && $-6.58\pm1.29$\\
&$\textrm{\textsc{elign}}_\textrm{self}$&$520.25\pm9.68$ && $510.18\pm25.71$ && $222.31\pm15.39$ && $-30.56\pm9.87$ && $-4.27\pm2.53$\\
\end{tabular}
}
\end{table*}

\begin{table*}[t]
\centering 
\caption{\label{tab:18_amb_scaled_dec_occupancy} The average test occupancy/collision count per step and standard errors of $\textit{decentralized}$ methods with different intrinsic rewards in $\textit{scaled}$ and $\textit{symmetry-breaking}$ settings. under partial and full observability. Higher scores are better for the occupancy metric ($\uparrow$), and lower scores are better for the collision metric ($\downarrow$). 
}
\resizebox{\textwidth}{!}{
\begin{tabular}{llrp{0.001pt}rp{0.001pt} rp{0.001pt}rp{0.001pt}r}
\multicolumn{2}{l}{}                & \multicolumn{3}{c}{\textbf{Cooperative}}                              & & \multicolumn{5}{c}{\textbf{Competitive}}                                                                            \\
\cmidrule{3-5}\cmidrule{7-11}
\multicolumn{2}{l}{Task (Agt \# vs. Adv \#) }                & \multicolumn{1}{c}{\textbf{Coop nav. (5v0)} $\uparrow$} && \multicolumn{1}{c}{\textbf{Hetero nav. (6v0)} $\uparrow$} && \multicolumn{1}{c}{\textbf{Phy decep. (4v2)} $\uparrow$}  && \multicolumn{1}{c}{\textbf{Pred-prey (4v4)} $\downarrow$} && \multicolumn{1}{c}{\textbf{Keep-away (4v4)} $\uparrow$} \\
\cmidrule{1-2}\cmidrule{3-3}\cmidrule{5-5}\cmidrule{7-7}\cmidrule{9-9}\cmidrule{11-11}
\multirow{4}{*}{\textbf{Partial observability}}  &$\textsc{sparse}$&$0.37\pm0.05$ && $0.38\pm0.03$ && $0.75\pm0.04$ && $0.13\pm0.03$ && $0.04\pm0.02$\\
&$\textrm{\textsc{curio}}_\textrm{self}$&$0.29\pm0.03$ && $0.43\pm0.02$ && $0.66\pm0.06$ && $0.05\pm0.02$ && $0.12\pm0.08$\\
&$\textrm{\textsc{curio}}_\textrm{team}$&$0.32\pm0.02$ && $0.34\pm0.01$ && $0.78\pm0.07$ && $0.05\pm0.02$ && $0.02\pm0.01$\\
&$\textrm{\textsc{elign}}_\textrm{self}$&$0.29	\pm 0.03$ && $0.39\pm0.02$ && $0.96\pm0.20$ && $0.04\pm0.03$ && $0.11\pm0.10$\\
&$\textrm{\textsc{elign}}_\textrm{team}$&$0.37\pm0.03$ && $0.40\pm0.03$ && $0.72\pm0.03$ && $0.07\pm0.03$ && $0.06\pm0.03$\\
&$\textrm{\textsc{elign}}_\textrm{adv}$&--- && --- && $0.63\pm0.09$ && $0.07\pm0.03$ && $0.15\pm0.10$\\

                         \hline
\multirow{2}{*}{\textbf{Full observability}}   
&$\textsc{sparse}$&$0.52\pm0.11$ && $0.43\pm0.09$ && $0.86\pm0.07$ && $0.19\pm0.02$ && $0.00\pm0.00$\\
&$\textrm{\textsc{curio}}_\textrm{self}$&$0.39\pm0.14$ && $0.54\pm0.00$ && $0.81\pm0.06$ && $0.14\pm0.04$ && $0.00\pm0.00$\\
&$\textrm{\textsc{elign}}_\textrm{self}$&$0.55\pm0.11$ && $0.55\pm0.00$ && $0.94\pm0.08$ && $0.12\pm0.03$ && $0.02\pm0.01$\\
\end{tabular}
}
\end{table*}

\begin{table*}[t]
\centering 
\caption{\label{tab:19_amb_scaled_dec_distance} The average test agent-to-target (agt-target) and agent-to-adversary (agt-adv) distances and standard errors of $\textit{decentralized}$ methods with different intrinsic rewards in $\textit{scaled}$ and $\textit{symmetry-breaking}$ settings under partial and full observability. Lower scores are better for agt-target ($\downarrow$), and higher scores are better for agt-adv ($\uparrow$).
}
\resizebox{\textwidth}{!}{
\begin{tabular}{llrp{0.001pt}rp{0.001pt} rp{0.001pt}rp{0.001pt}r}
\multicolumn{2}{l}{}                & \multicolumn{3}{c}{\textbf{Cooperative}}                              & & \multicolumn{5}{c}{\textbf{Competitive}}                                                                            \\
\cmidrule{3-5}\cmidrule{7-11}
\multicolumn{2}{l}{Task (Agt \# vs. Adv \#) }                & \multicolumn{1}{c}{\textbf{Coop nav. (5v0)} $\downarrow$} && \multicolumn{1}{c}{\textbf{Hetero nav. (6v0)} $\downarrow$} && \multicolumn{1}{c}{\textbf{Phy decep. (4v2)} $\downarrow$}  && \multicolumn{1}{c}{\textbf{Pred-prey (4v4)} $\uparrow$} && \multicolumn{1}{c}{\textbf{Keep-away (4v4)} $\downarrow$} \\
\cmidrule{1-2}\cmidrule{3-3}\cmidrule{5-5}\cmidrule{7-7}\cmidrule{9-9}\cmidrule{11-11}
\multirow{4}{*}{\textbf{Partial observability}}  &$\textsc{sparse}$&$0.36\pm0.01$ && $0.42\pm0.02$ && $0.35\pm0.01$ && $2.04\pm0.18$ && $3.10\pm0.29$\\
&$\textrm{\textsc{curio}}_\textrm{self}$&$0.50\pm0.04$ && $0.37\pm0.02$ && $0.38\pm0.03$ && $2.35\pm0.12$ && $2.70\pm0.35$\\
&$\textrm{\textsc{curio}}_\textrm{team}$&$0.43\pm0.03$ && $0.45\pm0.01$ && $0.35\pm0.02$ && $2.39\pm0.15$ && $3.37\pm0.19$\\
&$\textrm{\textsc{elign}}_\textrm{self}$&$0.52\pm0.05$ && $0.41\pm0.02$ && $0.37\pm0.04$ && $2.52\pm0.15$ && $3.22\pm0.42$\\
&$\textrm{\textsc{elign}}_\textrm{team}$&$0.42\pm0.03$ && $0.41\pm0.02$ && $0.37\pm0.01$ && $2.25\pm0.15$ && $3.00\pm0.33$\\
&$\textrm{\textsc{elign}}_\textrm{adv}$&--- && --- && $0.41\pm0.05$ && $2.27\pm0.16$ && $2.62\pm0.30$\\
                         \hline
\multirow{2}{*}{\textbf{Full observability}}   
&$\textsc{sparse}$&$0.29\pm0.07$ && $0.37\pm0.06$ && $0.26\pm0.01$ && $1.81\pm0.04$ && $3.69\pm0.05$\\
&$\textrm{\textsc{curio}}_\textrm{self}$&$0.42\pm0.11$ && $0.29\pm0.00$ && $0.26\pm0.02$ && $2.02\pm0.16$ && $3.63\pm0.12$\\
&$\textrm{\textsc{elign}}_\textrm{self}$&$0.26\pm0.05$ && $0.29\pm0.00$ && $0.27\pm0.01$ && $2.10\pm0.12$ && $3.24\pm0.26$\\
\end{tabular}
}
\end{table*}

\begin{table*}[t]
\centering 
\caption{\label{tab:8_central_reward_fo}We report the mean test episode extrinsic rewards and standard errors of $\textit{centralized}$ methods with different intrinsic rewards under full observability.
}
\resizebox{\textwidth}{!}{
\begin{tabular}{llrp{0.001pt}rp{0.001pt} rp{0.001pt}rp{0.001pt}r}
\multicolumn{2}{l}{}                & \multicolumn{3}{c}{\textbf{Cooperative}}                              & & \multicolumn{5}{c}{\textbf{Competitive}}                                                                            \\
\cmidrule{3-5}\cmidrule{7-11}
\multicolumn{2}{l}{Task (Agt \# vs. Adv \#) }                & \multicolumn{1}{c}{\textbf{Coop nav. (3v0)}} && \multicolumn{1}{c}{\textbf{Hetero nav. (4v0)}} && \multicolumn{1}{c}{\textbf{Phy decep. (2v1)}}  && \multicolumn{1}{c}{\textbf{Pred-prey (2v2)}} && \multicolumn{1}{c}{\textbf{Keep-away (2v2)}} \\
\cmidrule{1-2}\cmidrule{3-3}\cmidrule{5-5}\cmidrule{7-7}\cmidrule{9-9}\cmidrule{11-11}
\multirow{3}{*}{\textbf{Full observability}} &$\textsc{sparse}$&$106.02\pm20.95$ && $123.17\pm18.77$ && $130.90\pm6.59$ && $-1.90\pm1.61$ && $12.49\pm9.83$\\
&$\textrm{\textsc{curio}}_\textrm{self}$&$86.52\pm16.02$ && $108.84\pm6.89$ && $107.84\pm13.67$ && $-1.69\pm0.60$ && $23.70\pm12.95$\\
&$\textrm{\textsc{elign}}_\textrm{self}$&$120.47\pm12.26$ && $134.30\pm5.84$ && $105.74\pm9.72$ && $-2.37\pm1.39$ && $22.92\pm7.00$\\
\end{tabular}
}
\end{table*}

\begin{table*}[t]
\centering 
\caption{\label{tab:9_central_occupancy} The average test occupancy/collision count per step and standard errors of $\textit{centralized}$ methods with different intrinsic rewards under partial and full observability. Higher scores are better for the occupancy metric ($\uparrow$), and lower scores are better for the collision metric ($\downarrow$). 
}
\resizebox{\textwidth}{!}{
\begin{tabular}{llrp{0.001pt}rp{0.001pt} rp{0.001pt}rp{0.001pt}r}
\multicolumn{2}{l}{}                & \multicolumn{3}{c}{\textbf{Cooperative}}                              & & \multicolumn{5}{c}{\textbf{Competitive}}                                                                            \\
\cmidrule{3-5}\cmidrule{7-11}
\multicolumn{2}{l}{Task (Agt \# vs. Adv \#) }                & \multicolumn{1}{c}{\textbf{Coop nav. (3v0)} $\uparrow$} && \multicolumn{1}{c}{\textbf{Hetero nav. (4v0)} $\uparrow$} && \multicolumn{1}{c}{\textbf{Phy decep. (2v1)} $\uparrow$}  && \multicolumn{1}{c}{\textbf{Pred-prey (2v2)} $\downarrow$} && \multicolumn{1}{c}{\textbf{Keep-away (2v2)} $\uparrow$} \\
\cmidrule{1-2}\cmidrule{3-3}\cmidrule{5-5}\cmidrule{7-7}\cmidrule{9-9}\cmidrule{11-11}
\multirow{6}{*}{\textbf{Partial observability}}  &$\textsc{sparse}$&$0.29\pm0.10$ && $0.50\pm0.03$ && $0.94\pm0.06$ && $0.00\pm0.00$ && $0.36\pm0.12$\\
&$\textrm{\textsc{curio}}_\textrm{self}$&$0.28\pm0.09$ && $0.47\pm0.03$ && $0.94\pm0.03$ && $0.01\pm0.00$ && $0.17\pm0.10$\\
&$\textrm{\textsc{curio}}_\textrm{team}$&$0.33\pm0.10$ && $0.47\pm0.04$ && $0.92\pm0.01$ && $0.01\pm0.00$ && $0.08\pm0.03$\\
&$\textrm{\textsc{elign}}_\textrm{self}$&$0.21\pm0.10$ && $0.50\pm0.01$ && $0.92\pm0.02$ && $0.00\pm0.00$ && $0.25\pm0.08$\\
&$\textrm{\textsc{elign}}_\textrm{team}$&$0.23\pm0.09$ && $0.55\pm0.02$ && $0.90\pm0.07$ && $0.01\pm0.00$ && $0.24\pm0.11$\\
&$\textrm{\textsc{elign}}_\textrm{adv}$&--- && --- && $0.94\pm0.04$ && $0.00\pm0.00$ && $0.10\pm0.05$\\
                         \hline
\multirow{3}{*}{\textbf{Full observability}}   
&$\textsc{sparse}$&$0.34\pm0.10$ && $0.33\pm0.07$ && $0.88\pm0.04$ && $0.01\pm0.00$ && $0.26\pm0.11$\\
&$\textrm{\textsc{curio}}_\textrm{self}$&$0.30\pm0.07$ && $0.32\pm0.05$ && $0.82\pm0.02$ && $0.01\pm0.01$ && $0.33\pm0.16$\\
&$\textrm{\textsc{elign}}_\textrm{self}$&$0.30\pm0.11$ && $0.40\pm0.04$ && $0.88\pm0.05$ && $0.01\pm0.01$ && $0.30\pm0.07$\\
\end{tabular}
}
\end{table*}

\begin{table*}[t]
\centering 
\caption{\label{tab:10_central_distance} The average test agent-to-target (agt-target) and agent-to-adversary (agt-adv) distances and standard errors of $\textit{centralized}$ methods with different intrinsic rewards under partial and full observability. Lower scores are better for agt-target ($\downarrow$), and higher scores are better for agt-adv ($\uparrow$).
}
\resizebox{\textwidth}{!}{
\begin{tabular}{llrp{0.001pt}rp{0.001pt} rp{0.001pt}rp{0.001pt}r}
\multicolumn{2}{l}{}                & \multicolumn{3}{c}{\textbf{Cooperative}}                              & & \multicolumn{5}{c}{\textbf{Competitive}}                                                                            \\
\cmidrule{3-5}\cmidrule{7-11}
\multicolumn{2}{l}{Task (Agt \# vs. Adv \#) }                & \multicolumn{1}{c}{\textbf{Coop nav. (3v0)} $\downarrow$} && \multicolumn{1}{c}{\textbf{Hetero nav. (4v0)} $\downarrow$} && \multicolumn{1}{c}{\textbf{Phy decep. (2v1)} $\downarrow$}  && \multicolumn{1}{c}{\textbf{Pred-prey (2v2)} $\uparrow$} && \multicolumn{1}{c}{\textbf{Keep-away (2v2)} $\downarrow$} \\
\cmidrule{1-2}\cmidrule{3-3}\cmidrule{5-5}\cmidrule{7-7}\cmidrule{9-9}\cmidrule{11-11}
\multirow{6}{*}{\textbf{Partial observability}}  &$\textsc{sparse}$&$0.42\pm0.05$ && $0.29\pm0.02$ && $0.27\pm0.01$ && $1.54\pm0.02$ && $1.38\pm0.13$\\
&$\textrm{\textsc{curio}}_\textrm{self}$&$0.42\pm0.05$ && $0.29\pm0.01$ && $0.27\pm0.01$ && $1.46\pm0.05$ && $1.40\pm0.13$\\
&$\textrm{\textsc{curio}}_\textrm{team}$&$0.41\pm0.06$ && $0.29\pm0.02$ && $0.28\pm0.01$ && $1.49\pm0.04$ && $1.43\pm0.14$\\
&$\textrm{\textsc{elign}}_\textrm{self}$&$0.50\pm0.07$ && $0.29\pm0.01$ && $0.27\pm0.01$ && $1.60\pm0.04$ && $1.26\pm0.12$\\
&$\textrm{\textsc{elign}}_\textrm{team}$&$0.45\pm0.05$ && $0.27\pm0.01$ && $0.28\pm0.01$ && $1.52\pm0.04$ && $1.35\pm0.14$\\
&$\textrm{\textsc{elign}}_\textrm{adv}$&--- && --- && $0.28\pm0.01$ && $1.55\pm0.03$ && $1.45\pm0.10$\\
                         \hline
\multirow{3}{*}{\textbf{Full observability}}   
&$\textsc{sparse}$&$0.38\pm0.07$ && $0.34\pm0.04$ && $0.25\pm0.00$ && $1.59\pm0.06$ && $1.43\pm0.09$\\
&$\textrm{\textsc{curio}}_\textrm{self}$&$0.36\pm0.05$ && $0.32\pm0.02$ && $0.25\pm0.01$ && $1.53\pm0.09$ && $1.08\pm0.15$\\
&$\textrm{\textsc{elign}}_\textrm{self}$&$0.43\pm0.08$ && $0.30\pm0.02$ && $0.25\pm0.00$ && $1.51\pm0.08$ && $1.18\pm0.15$\\
\end{tabular}
}
\end{table*}

\begin{table*}[t]
\centering 
\caption{\label{tab:11_scaled_central_reward}We report the mean test episode extrinsic rewards and standard errors of $\textit{centralized}$ methods with different intrinsic rewards in $\textit{scaled}$ environments. 
}
\resizebox{\textwidth}{!}{
\begin{tabular}[width=\linewidth]{llrp{0.001pt}rp{0.001pt} rp{0.001pt}rp{0.001pt}r}
\multicolumn{2}{l}{}                & \multicolumn{3}{c}{\textbf{Cooperative}}                              & & \multicolumn{5}{c}{\textbf{Competitive}}                                                                            \\
\cmidrule{3-5}\cmidrule{7-11}
\multicolumn{2}{l}{Task (Agt \# vs. Adv \#) }                & \multicolumn{1}{c}{\textbf{Coop nav. (5v0)}} && \multicolumn{1}{c}{\textbf{Hetero nav. (6v0)}} && \multicolumn{1}{c}{\textbf{Phy decep. (4v2)}}  && \multicolumn{1}{c}{\textbf{Pred-prey (4v4)}} && \multicolumn{1}{c}{\textbf{Keep-away (4v4)}} \\
\cmidrule{1-2}\cmidrule{3-3}\cmidrule{5-5}\cmidrule{7-7}\cmidrule{9-9}\cmidrule{11-11}
\multirow{5}{*}{\textbf{Partial observability}} &$\textsc{sparse}$&$100.63\pm19.36$ && $346.16\pm18.95$ && $-38.99\pm16.18$ && $-17.33\pm4.29$ && $-2.50\pm2.64$\\
&$\textrm{\textsc{elign}}_\textrm{self}$&$112.15\pm19.69$ && $375.21\pm26.10$ && $13.71\pm29.53$ && $-20.12\pm1.42$ && $-4.68\pm1.21$\\
&$\textrm{\textsc{elign}}_\textrm{team}$&$97.93\pm25.23$ && $372.41\pm44.28$ && $60.07\pm13.26$ && $-27.87\pm0.99$ && $1.72\pm3.79$\\
&$\textrm{\textsc{elign}}_\textrm{adv}$&--- && --- && $21.67\pm48.17$ && $-17.68\pm5.59$ && $-4.92\pm1.81$\\
                               \hline
\multirow{2}{*}{\textbf{Full observability}} &$\textsc{sparse}$&$50.60\pm13.10$ && $153.76\pm19.81$ && $97.32\pm17.95$ && $-38.25\pm5.06$ && $-3.39\pm2.77$\\
&$\textrm{\textsc{elign}}_\textrm{self}$&$186.55\pm53.15$ && $127.97\pm13.02$ && $103.46\pm28.91$ && $-23.29\pm5.00$ && $-4.90\pm0.67$\\
\end{tabular}
}
\end{table*}

\begin{table*}[t]
\centering 
\caption{\label{tab:12_scaled_central_occupancy} The average test occupancy/collision count per step and standard errors of $\textit{centralized}$ methods with different intrinsic rewards in \textit{scaled} environments under partial and full observability. Higher scores are better for the occupancy metric ($\uparrow$), and lower scores are better for the collision metric ($\downarrow$). 
}
\resizebox{\textwidth}{!}{
\begin{tabular}{llrp{0.001pt}rp{0.001pt} rp{0.001pt}rp{0.001pt}r}
\multicolumn{2}{l}{}                & \multicolumn{3}{c}{\textbf{Cooperative}}                              & & \multicolumn{5}{c}{\textbf{Competitive}}                                                                            \\
\cmidrule{3-5}\cmidrule{7-11}
\multicolumn{2}{l}{Task (Agt \# vs. Adv \#) }                & \multicolumn{1}{c}{\textbf{Coop nav. (5v0)} $\uparrow$} && \multicolumn{1}{c}{\textbf{Hetero nav. (6v0)} $\uparrow$} && \multicolumn{1}{c}{\textbf{Phy decep. (4v2)} $\uparrow$}  && \multicolumn{1}{c}{\textbf{Pred-prey (4v4)} $\downarrow$} && \multicolumn{1}{c}{\textbf{Keep-away (4v4)} $\uparrow$} \\
\cmidrule{1-2}\cmidrule{3-3}\cmidrule{5-5}\cmidrule{7-7}\cmidrule{9-9}\cmidrule{11-11}
\multirow{4}{*}{\textbf{Partial observability}}  &$\textsc{sparse}$&$0.11\pm0.02$ && $0.29\pm0.06$ && $0.56\pm0.05$ && $0.06\pm0.02$ && $0.07\pm0.02$\\
&$\textrm{\textsc{elign}}_\textrm{self}$&$0.23\pm0.09$ && $0.33\pm0.04$ && $0.56\pm0.06$ && $0.08\pm0.00$ && $0.05\pm0.00$\\
&$\textrm{\textsc{elign}}_\textrm{team}$&$0.27\pm0.10$ && $0.33\pm0.05$ && $0.50\pm0.08$ && $0.09\pm0.00$ && $0.09\pm0.03$\\
&$\textrm{\textsc{elign}}_\textrm{adv}$&--- && --- && $0.60\pm0.09$ && $0.05\pm0.02$ && $0.05\pm0.01$\\
                         \hline
\multirow{2}{*}{\textbf{Full observability}}   
&$\textsc{sparse}$&$0.10\pm0.04$ && $0.16\pm0.04$ && $0.50\pm0.03$ && $0.12\pm0.01$ && $0.06\pm0.01$\\
&$\textrm{\textsc{elign}}_\textrm{self}$&$0.16\pm0.09$ && $0.11\pm0.00$ && $0.55\pm0.02$ && $0.10\pm0.02$ && $0.04\pm0.01$\\
\end{tabular}
}
\end{table*}

\begin{table*}[t]
\centering 
\caption{\label{tab:13_scaled_central_distance} The average test agent-to-target (agt-target) and agent-to-adversary (agt-adv) distances and standard errors of $\textit{centralized}$ methods with different intrinsic rewards in \textit{scaled} environments under partial and full observability. Lower scores are better for agt-target ($\downarrow$), and higher scores are better for agt-adv ($\uparrow$).
}
\resizebox{\textwidth}{!}{
\begin{tabular}{llrp{0.001pt}rp{0.001pt} rp{0.001pt}rp{0.001pt}r}
\multicolumn{2}{l}{}                & \multicolumn{3}{c}{\textbf{Cooperative}}                              & & \multicolumn{5}{c}{\textbf{Competitive}}                                                                            \\
\cmidrule{3-5}\cmidrule{7-11}
\multicolumn{2}{l}{Task (Agt \# vs. Adv \#) }                & \multicolumn{1}{c}{\textbf{Coop nav. (5v0)} $\downarrow$} && \multicolumn{1}{c}{\textbf{Hetero nav. (6v0)} $\downarrow$} && \multicolumn{1}{c}{\textbf{Phy decep. (4v2)} $\downarrow$}  && \multicolumn{1}{c}{\textbf{Pred-prey (4v4)} $\uparrow$} && \multicolumn{1}{c}{\textbf{Keep-away (4v4)} $\downarrow$} \\
\cmidrule{1-2}\cmidrule{3-3}\cmidrule{5-5}\cmidrule{7-7}\cmidrule{9-9}\cmidrule{11-11}
\multirow{4}{*}{\textbf{Partial observability}}  &$\textsc{sparse}$&$0.33\pm0.01$ && $0.29\pm0.02$ && $0.34\pm0.02$ && $2.27\pm0.08$ && $3.12\pm0.18$\\
&$\textrm{\textsc{elign}}_\textrm{self}$&$0.32\pm0.04$ && $0.28\pm0.01$ && $0.36\pm0.02$ && $2.32\pm0.09$ && $3.25\pm0.05$\\
&$\textrm{\textsc{elign}}_\textrm{team}$&$0.30\pm0.04$ && $0.28\pm0.01$ && $0.37\pm0.03$ && $2.29\pm0.08$ && $3.01\pm0.20$\\
&$\textrm{\textsc{elign}}_\textrm{adv}$&--- && --- && $0.33\pm0.04$ && $2.44\pm0.08$ && $3.23\pm0.13$\\

                         \hline
\multirow{2}{*}{\textbf{Full observability}}   
&$\textsc{sparse}$&$0.37\pm0.03$ && $0.37\pm0.02$ && $0.29\pm0.02$ && $1.98\pm0.07$ && $3.13\pm0.17$\\
&$\textrm{\textsc{elign}}_\textrm{self}$&$0.36\pm0.04$ && $0.39\pm0.00$ && $0.27\pm0.01$ && $1.98\pm0.08$ && $3.25\pm0.12$\\
\end{tabular}
}
\end{table*}

\begin{table*}[t]
\centering 
\caption{\label{tab:20_zero_shot_scaled_dec_reward} We sample agents from different \textit{decentralized} training runs and evaluate their zero-shot performance in \textit{scaled} environments under partial observability. We report the mean test episode extrinsic rewards and standard errors of decentralized methods with different intrinsic rewards.
}
\resizebox{\textwidth}{!}{
\begin{tabular}{llrp{0.001pt}rp{0.001pt} rp{0.001pt}rp{0.001pt}r}
\multicolumn{2}{l}{}                & \multicolumn{3}{c}{\textbf{Cooperative}}                              & & \multicolumn{5}{c}{\textbf{Competitive}}                                                                            \\
\cmidrule{3-5}\cmidrule{7-11}
\multicolumn{2}{l}{Task (Agt \# vs. Adv \#) }                & \multicolumn{1}{c}{\textbf{Coop nav. (5v0)}} && \multicolumn{1}{c}{\textbf{Hetero nav. (6v0)}} && \multicolumn{1}{c}{\textbf{Phy decep. (4v2)}}  && \multicolumn{1}{c}{\textbf{Pred-prey (4v4)}} && \multicolumn{1}{c}{\textbf{Keep-away (4v4)}} \\
\cmidrule{1-2}\cmidrule{3-3}\cmidrule{5-5}\cmidrule{7-7}\cmidrule{9-9}\cmidrule{11-11}
\multirow{4}{*}{\textbf{Partial observability}} &$\textsc{sparse}$&$434.68\pm6.42$ && $561.16\pm31.63$ && $128.64\pm17.31$ && $-32.12\pm3.63$ && $-2.80\pm2.91$\\
&$\textrm{\textsc{elign}}_\textrm{self}$&$471.07\pm5.00$ && $676.01\pm16.53$ && $248.16\pm6.62$ && $-16.77\pm2.25$ && $-5.03\pm1.06$\\
&$\textrm{\textsc{elign}}_\textrm{team}$&$511.97\pm6.95$ && $699.56\pm11.64$ && $190.06\pm29.10$ && $-19.40\pm2.86$ && $-3.10\pm3.09$\\
&$\textrm{\textsc{elign}}_\textrm{adv}$&--- && --- && $228.53\pm25.03$ && $-31.03\pm3.13$ && $27.24\pm4.48$\\
\end{tabular}
}
\end{table*}

\begin{table*}[t]
\centering 
\caption{\label{tab:21_zero_shot_scaled_dec_distance} We sample agents from different \textit{decentralized} training runs and evaluate their zero-shot performance in \textit{scaled} environments under partial observability. We report the average test agent-to-target (agt-target) and agent-to-adversary (agt-adv) distances and standard errors of $\textit{decentralized}$ methods with different intrinsic rewards. Lower scores are better for agt-target ($\downarrow$), and higher scores are better for agt-adv ($\uparrow$).
}
\resizebox{\textwidth}{!}{
\begin{tabular}{llrp{0.001pt}rp{0.001pt} rp{0.001pt}rp{0.001pt}r}
\multicolumn{2}{l}{}                & \multicolumn{3}{c}{\textbf{Cooperative}}                              & & \multicolumn{5}{c}{\textbf{Competitive}}                                                                            \\
\cmidrule{3-5}\cmidrule{7-11}
\multicolumn{2}{l}{Task (Agt \# vs. Adv \#) }                & \multicolumn{1}{c}{\textbf{Coop nav. (5v0)} $\downarrow$} && \multicolumn{1}{c}{\textbf{Hetero nav. (6v0)} $\downarrow$} && \multicolumn{1}{c}{\textbf{Phy decep. (4v2)} $\downarrow$}  && \multicolumn{1}{c}{\textbf{Pred-prey (4v4)} $\uparrow$} && \multicolumn{1}{c}{\textbf{Keep-away (4v4)} $\downarrow$} \\
\cmidrule{1-2}\cmidrule{3-3}\cmidrule{5-5}\cmidrule{7-7}\cmidrule{9-9}\cmidrule{11-11}
\multirow{4}{*}{\textbf{Partial observability}}  &$\textsc{sparse}$&$0.22\pm0.00$ && $0.23\pm0.00$ && $0.23\pm0.00$ && $1.93\pm0.00$ && $3.16\pm0.01$\\
&$\textrm{\textsc{elign}}_\textrm{self}$&$0.19\pm0.00$ && $0.20\pm0.00$ && $0.17\pm0.00$ && $2.33\pm0.00$ && $3.31\pm0.01$\\
&$\textrm{\textsc{elign}}_\textrm{team}$&$0.43\pm0.01$ && $0.19\pm0.00$ && $0.24\pm0.00$ && $2.04\pm0.01$ && $3.15\pm0.01$\\
&$\textrm{\textsc{elign}}_\textrm{adv}$&--- && --- && $0.21\pm0.00$ && $2.11\pm0.01$ && $2.31\pm0.01$\\
\end{tabular}
}
\end{table*}

\subsection{Model architecture and hyperparameters}
\begin{table*}[t]
\centering 
\caption{\label{tab:hyperparams} Model and training hyperparameters}
\begin{tabular}{lll} \toprule
Parameter & Multi-agent particle & Google Research football  \\ \midrule
SAC actor model architecture & FC layers [128,128] & FC layers [256,256] \\
SAC critic model architecture & FC layers [128,128] & FC layers [256,256] \\
World model architecture & FC layers [128,128] & FC layers [128,128] \\
Replay buffer size & 1,000,000 & 1,000,000\\ 
Batch size & 1,024 & 256\\
Actor learning rate & 0.001  & 0.0003\\
Critic learning rate & 0.001 & 0.0003 \\
Discount factor gamma & 0.95 & 0.99 \\
SAC soft update coefficient & 0.01 & 0.005 \\
SAC policy entropy regularization coefficient & 0.1  & 1.0 (initial)\\
\bottomrule
\end{tabular}
\end{table*}
Table~\ref{tab:hyperparams} presents the model architecture and hyperparameters used to train the algorithms in the multi-agent particle and Google Research football environments.

\end{document}